%% file: GenAdapQMMM.tex
\documentclass[12pt]{amsart}

\rmfamily{\fontsize{12pt}{\baselineskip}\selectfont}
\usepackage{graphicx}
\usepackage{amsfonts,amsmath,amssymb,amsbsy,amsthm}
\usepackage{bm}
\usepackage{color}
\usepackage{xcolor}
\usepackage{float}
\usepackage{hyperref}
\usepackage{mathrsfs}
\usepackage{longtable}
\usepackage[top=2.5cm,bottom=2.0cm,left=2.0cm,right=2.0cm]{geometry}
\usepackage{epstopdf}
\usepackage{stmaryrd}
\usepackage[normalem]{ulem}
\usepackage{scalerel}
\usepackage{enumitem}
\usepackage{algorithm}
\usepackage{subfig}
\usepackage{algorithm, algorithmicx}
\usepackage[noend]{algpseudocode}
\usepackage{algorithm, algorithmicx}
\usepackage{lineno}
\usepackage{enumitem}
\renewcommand{\algorithmiccomment}[1]{\bgroup\hfill//~#1\egroup}
\usepackage{caption}

\usepackage{environments}
\numberwithin{theorem}{section}
\numberwithin{equation}{section}

\input{notation}

\def\Lhom{\Lambda^{\rm hom}}

\definecolor{yscol}{HTML}{6622AA}

\begin{document}
\title[Adaptivity for QM/MM Models]{A Posteriori Error Estimate and Adaptivity \\
for QM/MM Models of Crystalline Defects}

\author{Yangshuai Wang}
\address{University of British Columbia, 1984 Mathematics Road, Vancouver, BC, Canada.}
\email{yswang2021@math.ubc.ca}
\author{James R. Kermode}
\address{Warwick Centre for Predictive Modelling, School of Engineering, University of Warwick, Coventry, CV4 7AL, United Kingdom.}
\email{J.R.Kermode@warwick.ac.uk}
\author{Christoph Ortner}
\address{University of British Columbia, 1984 Mathematics Road, Vancouver, BC, Canada.}
\email{ortner@math.ubc.ca}
\author{Lei Zhang}
\address{School of Mathematical Sciences, Institute of Natural Sciences and MOE-LSC, Shanghai Jiao Tong University, Shanghai 200240, China.}
\email{lzhang2012@sjtu.edu.cn}

\date{\today}




\begin{abstract}
Hybrid quantum/molecular mechanics (QM/MM) models play a pivotal role in molecular simulations. These models provide a balance between accuracy, surpassing pure MM models, and computational efficiency, offering advantages over pure QM models. Adaptive approaches have been developed to further improve this balance by allowing on-the-fly selection of the QM and MM subsystems as necessary. 
We propose a novel and robust adaptive QM/MM method for practical material defect simulations. To ensure mathematical consistency with the QM reference model, we employ machine-learning interatomic potentials (MLIPs) as the MM models~\cite{2021-qmmm3, grigorev2023calculation}.
Our adaptive QM/MM method utilizes a residual-based error estimator that provides both upper and lower bounds for the approximation error, thus indicating its reliability and efficiency. Furthermore, we introduce a novel adaptive algorithm capable of anisotropically updating the QM/MM partitions. This update is based on the proposed residual-based error estimator and involves solving a free interface motion problem, which is efficiently achieved using the fast marching method. We demonstrate the robustness of our approach via numerical tests on a wide range of crystalline defects. 
\end{abstract}

\maketitle


\input{notation}
\section{Introduction}
\label{sec:intro}

Quantum mechanics and molecular mechanics (QM/MM) coupling methods have emerged as indispensable tools for simulating large molecular systems in both materials science and biology~\cite{bernstein09, csanyi04, gao02, kermode08, ogata01, zhang12}. The essence of this methodology lies in the partition of the computational domain into QM and MM regions, where the region of primary interest is described by a QM model embedded in an ambient environment represented by a MM model. The goal of QM/MM coupling methods is to achieve (near-)QM accuracy at (near-)MM computational cost, to make quantitative large-scale atomistic simulations viable.

An important question in the field of QM/MM coupling methods concerns the optimal assignment of atoms to either QM or MM subsystems in order to balance accuracy and computational cost. To address this issue, adaptive QM/MM methods have been developed by providing an on-the-fly partition of QM/MM subsystems based on error estimators during simulations. 

Adaptive QM/MM methods have been proposed for various applications such as the analysis of molecular fragments in macromolecules, monitoring of molecules entering/leaving binding sites and tracking proton transfer through the Grotthuss mechanism (as discussed in \cite{duster17} and references therein). However, the majority of adaptive QM/MM methods developed thus far are for solute-solvent systems and rely on heuristic criteria. Further information can be found in \cite{boereboom2016, Glukhova:2014, heyden2007, kerdcharoen1996, waller2014, watanabe2014, ZYang2020}. In the context of materials featuring defects, adaptive computations are often guided by the distance to these defects~\cite{csanyi04, kermode08}. Similar concepts can also be observed in the quasi-continuum methods for density functional theory~\cite{Gavini:2007, Ponga:2016}.

Adaptive QM/MM methods normally rely on empirical error estimators. 
By contrast, Chen et al. \cite{CMAME} first introduced mathematically rigorous {\it a posteriori} error estimators for the QM/MM model residual inspired by classical adaptive finite element methods \cite{Dorfler:1996, Verfurth:1996a}. Their approach employed a weighted $\ell^2$-norm on the {\it residual} forces, providing an upper bound for the approximation error. However, it fell short in providing a lower bound, which is essential for ensuring efficiency of adaptive schemes. To overcome this limitation, Wang et al. \cite{wang2020posteriori} proposed a reliable and efficient {\it a posteriori} error estimator by connecting the natural dual norm of the residual with solving an auxiliary Poisson equation. They developed an inner-outer adaptive strategy with an outer adaptive algorithm for selecting QM and MM regions and an inner algorithm for computing the estimators with the desired accuracy. However, the inner algorithm necessitated a finite element mesh, which introduced additional errors requiring careful handling, not to mention additional algorithmic complexity. More importantly, this work was primarily centered on energy-mixing schemes for simple point defects and the adaptive algorithm only adjusts the radius of QM and MM sub-regions, limiting the ability to capture significant anisotropy in the defect core, elastic field, or defect nucleation observed in practical material simulations. 

The purpose of the present work is to develop a more practical adaptive QM/MM method for material defect simulations, while maintaining the rigourous approach of \cite{CMAME, wang2020posteriori}. To ensure {\it consistency} of the QM/MM scheme and improve computational efficiency, we employ state-of-art machine-learning interatomic potentials (MLIPs) as the MM models~\cite{2021-qmmm3, grigorev2023calculation}. 
Next, we propose a practical and flexible approach to obtain the error estimator, essentially replacing the PDE operator from \cite{wang2020posteriori} with a generalization of the graph-Laplacian \cite{2016-precon1}. Algorithmically, this approach fits much better into the setting of atomistic modeling. A practical error estimator is further developed by (i) truncating to a finite computational domain and (ii) facilitating the QM force constant to give a linear approximation of the residual force in the MM region.
Moreover, to evolve the QM/MM partitions anisotropically rather than only adjusting the radius, a free interface motion problem (i.e., Eikonal equation \cite{zhao2005fast}) is formulated and solved using the fast marching method \cite{chopp2001some, zhao2005fast}, where the {\it practical} error estimator is regarded as the extending speed. We develop a novel strategy to assign atoms to QM or MM subsystems based on the solution of the corresponding Eikonal equation. 

We test our algorithm by performing adaptive computations for three common defect types: edge dislocations, in-plane cracks, and di-interstitials. Our findings reveal that the practical error estimator we introduce attains convergence rates comparable to those of the approximation error, offering substantial computational cost reductions when employing a realistic electronic structure model. The adaptive algorithm showcases robustness by eliminating the need for user {\it a priori} input, thereby aligning with our objective of achieving a fully adaptive QM/MM scheme. The analysis and adaptive algorithm presented in this paper demonstrates a considerable degree of independence from the underlying approximation scheme, thereby rendering the proposed framework widely applicable to various coarse-graining or multiscale methods. As a proof of concept, we will focus solely on geometry equilibration problems (statics). 

\subsection*{Outline}
This paper is organized as follows: Section~\ref{sec:QMMM} introduces the variational formulation for defect equilibration and the QM/MM coupling methods we use. Section~\ref{sec:errest} outlines the construction of our novel {\it a posteriori} error estimator, which provides upper and lower bounds for the approximation error, along with practical approximations to enhance its implementation. Section~\ref{sec:alg_numerics} presents our adaptive QM/MM algorithm, incorporating a free interface problem to dynamically update QM/MM partitions using the practical error estimator. We showcase our findings with numerical examples, validating the efficacy of our adaptive algorithm. Section~\ref{sec:conclusion} concludes our key findings and outlines future research directions. Appendices provide supplementary information for interested readers.

\subsection*{Notation}
We use the symbol $\langle\cdot,\cdot\rangle$ to denote the duality
pairing between a Banach space and its dual space. The symbol $|\cdot|$ normally
denotes the Euclidean or Frobenius norm, while $\|\cdot\|$ denotes an operator
norm.
For the sake of brevity of notation, we will denote $A\backslash\{a\}$ by
$A\backslash a$, and $\{b-a~\vert ~b\in A\}$ by $A-a$.
For $E \in C^2(X)$, the first and second variations are denoted by
$\<\delta E(u), v\>$ and $\<\delta^2 E(u) v, w\>$ for $u,v,w\in X$.
For a finite set $A$, we will use $\#A$ to denote the cardinality of $A$.
The closed ball with radius $r$ and center $x$ is denoted by $B_r(x)$, or $B_r$ if the center is the origin.
We use the standard definitions and notations $L^p$, $W^{k,p}$, $H^k$ for Lebesgue and Sobolev spaces. In addition we define the homogeneous Sobolev spaces
$
	\dot{H}^k(\Omega) := \big\{ f \in H^k_{\rm loc}(\Omega) \,|\,
										\nabla^k f \in L^2(\Omega) \big\}.
$

\section{QM/MM Coupling for Crystalline Defects}
\label{sec:QMMM}

\subsection{Variational model for crystalline defects}
\label{sec:sub:var}

A rigorous framework for modelling the geometric equilibration of crystalline defects has been developed in~\cite{chen19, Ehrlacher16}. These works formulate the equilibration of a single crystalline defect as a variational problem in a discrete energy space and establish qualitatively sharp far-field decay estimates for the corresponding equilibrium configuration. This analytical foundation will serve as the basis for our {\it a posteriori} error estimates and the corresponding adaptive algorithm.

\subsubsection{Displacement space}
Given $d \in \{ 2, 3 \}$, a homogeneous crystal reference configuration is given by the Bravais lattice
$\Lhom=\mathsf{A}\Z^d$, for some non-singular matrix $\mathsf{A} \in \mathbb{R}^{d \times d}$. A reference lattice with a single defect is denoted by $\L \subset \R^d$. For the sake of simplicity we admit only single-species Bravais lattices. There are no conceptual obstacles to generalising this work to multi-lattices~\cite{olson2019theoretical}, however, the notational and technical details become more involved. 

A deformed configuration of the infinite lattice $\L$ is a map $y: \L\rightarrow\R^d$. We can decompose the configuration $y$ into
\begin{eqnarray}\label{y-u}
    y(\ell) = \ell + u_0(\ell) + u(\ell) = y_0(\ell) + u(\ell),
\end{eqnarray}
where $u_0: \L\rightarrow\R^d$ is a {\it far-field predictor} solving a continuum linearised elasticity (CLE) equation~\cite{Ehrlacher16} enforcing the presence of defect and $u: \L\rightarrow\R^d$ is a {\it core corrector}. For point defects we simply take $u_0(\ell)=0~\forall~\ell\in\L$. The derivation of $u_0$ for straight dislocations and cracks are reviewed in the Appendix \ref{sec:appendixU0}.

The set of admissible atomic configurations is
\begin{align*}
\Adm_{0}(\L) &:= \bigcup_{\mathfrak{m}>0} \Adm_{\mathfrak{m}}(\L)
\qquad \text{with} \\
\Adm_{\mathfrak{m}}(\L) &:= \left\{ y:\L \rightarrow \R^{d}, ~
|y(\ell)-y(m)| > \mathfrak{m} |\ell-m|
\quad\forall~  \ell, m \in \L \right\},
\end{align*}
where the parameter $\mathfrak{m}>0$ prevents the accumulation of atoms.

For $\ell\in\L$ and $\rho\in\L-\ell$, we define the finite difference
$D_\rho u(\ell) := u(\ell+\rho) - u(\ell)$. For a subset $\Rl \subset \Lambda-\ell$, we
define $D_{\Rl} u(\ell) := (D_\rho u(\ell))_{\rho\in\Rl}$, and we consider $Du(\ell) := D_{\Lambda-\ell} u(\ell)$ to be a finite-difference stencil with infinite range.
For a stencil $Du(\ell)$, we define the stencil norms
\begin{align}\label{eq: nn norm}
\big|Du(\ell)\big|_{\mathcal{N}} := \bigg( \sum_{\rho\in \mathcal{N}(\ell) - \ell} \big|D_\rho u(\ell)\big|^2 \bigg)^{1/2}  
\quad{\rm and}\quad
\|Du\|_{\ell^2_{\mathcal{N}}} := \bigg( \sum_{\ell \in \L} |Du(\ell)|_{\mathcal{N}}^2 \bigg)^{1/2},
\end{align}
where $\mathcal{N}(\ell)$ is the set containing nearest neighbours of site $\ell$,
\begin{align}\label{def1:Nl}
\mathcal{N}(\ell) :=& \left\{ \, m \in \L \setminus \ell \,~\Big|~\, \exists \, a \in \mathbb{R}^{d} \text{ s.t. }
|a - \ell| = |a - m| \leq |a - k| \quad \forall \, k \in \L \, \right\}. \,\,
\end{align}
We can then define the corresponding functional space of finite-energy displacements
\begin{align}\label{space:UsH}
\UsH(\L) := \big\{u:\L\rightarrow\mathbb{R}^{d} ~\big\lvert~ \|Du\|_{\ell^2_{\mathcal{N}}(\L)}<\infty \big\}
\end{align}
with the associated semi-norm  $\|u\|_{\UsH} := \|Du\|_{\ell^2_{\mathcal{N}}}$.
Then the associated class of admissible displacements is given by
\begin{eqnarray*}
	\Admu(\L):= \big\{ u\in\UsH(\L) ~:~ y_0 +u\in\Adm_0(\L) \big\}.
\end{eqnarray*}

\subsubsection{The QM site potential}
We consider the site potential to be a collection of mappings $V_{\ell}:(\R^d)^{\L-\ell}\rightarrow\R$, which represent the energy distributed to each atomic site.
For technical reasons we make the following assumptions on the regularity and locality of the site potentials, which has been justified for some basic quantum mechanic models \cite{chen18,chen16,chen19tb, co2020}, but we emphasize that it is not a universally valid assumption.

\begin{itemize}
	\label{as:SE:pr}
	\item[\assERL]
	{\it Regularity and locality:}
	For all $\ell \in \L$, $V_{\ell}\big(Du(\ell)\big)$ possesses partial derivatives up to the third order. For $j=1,2,3$, there exist constants $C_j$ and $\eta_j$ such that 
	\begin{eqnarray}
	\label{eq:Vloc}
	\big|V_{\ell,{\bm \rho}}\big(Du(\ell)\big)\big|  \leq
	C_j \exp\Big(-\eta_j\sum^j_{l=1}|{\bm \rho}_l|\Big)
	\end{eqnarray} 
	for all $\ell \in \L$ and ${\bm \rho} \in (\L - \ell)^{j}$. 
\end{itemize}
We refer to \cite[\S 2.3 and \S 4]{chen19} for a justification and discussion of this assumption. 


If the reference configuration $\L$ is a homogeneous lattice, $\L = \Lhom$, then the site potential does not depend on site $\ell\in\Lhom$ due to the translation invariance. In this case, 
we will denote the site potential by $\Vhom:(\R^d)^{\Lhom\setminus 0}\rightarrow\R$.
for the homogeneous lattice. Although the site potentials are defined on infinite stencils $(\R^d)^{\L-\ell}$, the setting also applies to finite systems or to finite range interactions. In particular, we denote by $V_{\ell}^{\Omega}$ the site potential of a finite system with the reference configuration lying in $\L\cap\Omega$.

\subsubsection{Equilibration of crystalline defects}
%
\def\Rdef{R^{\rm def}}
\def\Rg{\mathcal{R}}
\def\Rgnn{\mathcal{N}}
\def\rcut{R_{\rm c}}
\def\Ddef{D^{\rm def}}
\def\Ldef{\L^{\rm def}}
\def\Rcore{R_{\rm DEF}}
\def\Adm{{\rm Adm}}
\def\E{\mathcal{E}}
\def\L{\Lambda}
\def\UsH{{\mathscr{U}}^{1,2}}
\def\Usz{{\mathscr{U}^{\rm c}}}
\def\DD{{\sf D}}
\def\ee{{\sf e}}
Let the site potential $V_{\ell}$ satisfy the assumptions \assERL. The energy-difference functional is then given by 
\begin{eqnarray}\label{energy-difference}
\E(u) := \sum_{\ell\in\Lambda}\Big(V_{\ell}\big(Du_0(\ell)+Du(\ell)\big)-V_{\ell}\big(Du_0(\ell)\big)\Big).
\end{eqnarray}
It was shown in~\cite[Theorem 2.1]{chen19} that after an elementary renormalisation, \eqref{energy-difference} is well-defined on the admissible displacements set $\Admu(\L)$, and that it is $(\n-1)$-times continuously differentiable with respect to the $\|\cdot\|_{\UsH}$ norm.

Instead of the energy minimization problem~\cite{2021-qmmm3, wang2020posteriori}, we will focus on the (formally equivalent) force equilibrium formulation, that is, 
\begin{eqnarray}
\label{eq:problem-force}
{\rm Find}~ \bar{u} \in \Admu(\L), ~~ {\rm s.t.} \quad
\F_{\ell}(\bar{u}) = 0, \qquad \forall~\ell\in\Lambda ,
\end{eqnarray}
where $\F_{\ell}:= -\nabla_{\ell} \E(u)$ represents the force on the atomic site $\ell$. For $u^{\rm h}\in\Admu(\Lhom)$ on a homogeneous lattice $\Lhom$, the force on each atomic site $\ell$ satisfies $\F_{\ell}(u^{\rm h})=\F^{\rm h}\big(u^{\rm h}(\cdot-\ell)\big)$ with some homogeneous force $\F^{\rm h}$ that does not depend on site $\ell$.

\subsection{QM/MM Coupling}
\label{sec:sub:qmmm}

In this section, we describe the QM/MM models utilized in our study. Our approach is inspired by~\cite{chen17, 2021-qmmm3}, tailored to better align with the specific context of the present work.

\subsubsection{Domain decomposition}

We first decompose the reference configuration $\L$ into three disjoint sets,
$\Lambda = \LQM\cup \LMM\cup \LFF$, where $\LQM$ and $\LMM$ denote the QM and MM regions, respectively, and $\LFF$ represents the far-field region where atom positions are frozen. This yields the approximated admissible set
\begin{eqnarray}
\label{e-mix-space}
\Admu^{\rm H}(\L):= \big\{ u\in\Us^{1,2}(\L) ~:~ y_0+u\in\Adm_0(\L) ~~{\rm and}~~u=0~{\rm in}~\Lambda^{\rm FF} \big\}.
\end{eqnarray}
Moreover, we define a buffer region $\Lbuf\subset\LMM$ surrounding $\LQM$ such that all atoms in $\Lbuf\cup\LQM$ are involved in the evaluation of the site energies defined on $\LQM$.

\begin{figure}[!htb]
        \centering
        \includegraphics[height=5.5cm]{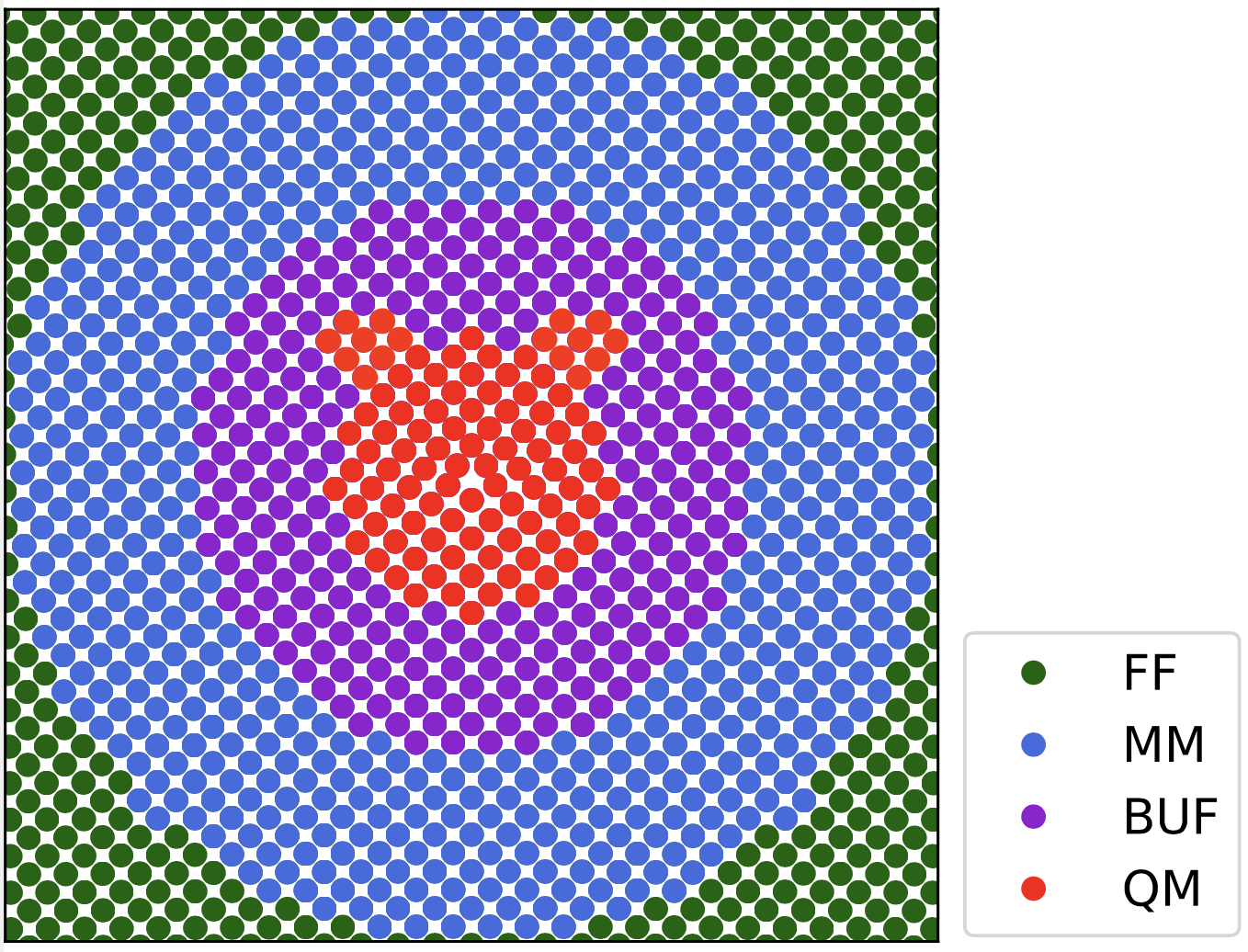}
        \caption{Decomposition of (001)[100] edge dislocation in W into QM, MM, buffer (BUF), and far-field (FF) regions. 
		}
        \label{fig:decomposition}
\end{figure}

In contrast to the method proposed in \cite{CMAME, wang2020posteriori}, which employed approximately spherical regions centered at the defect core for domain partitioning, we introduce a more versatile domain decomposition scheme in our study. Rather than relying on region radii as model parameters, we utilize the number of atoms within the QM, buffer, and MM regions, denoted respectively by $N_{\rm QM}:=\#\LQM$, $N_{\rm BUF}:=\#\Lbuf$, and $N_{\rm MM}:=\#\LMM$. Figure \ref{fig:decomposition} presents a typical QM/MM decomposition for a (001)[100] edge dislocation in Tungsten (W) with non-spherical subdomains.

\subsubsection{Specification of MM model}

To represent interatomic interactions at a distance from the defect core in the MM region, it is desirable to construct the MM site energy (force) that is computationally efficient, accurately captures the underlying physical behavior, and is ``compatible'' with the QM model. These requirements have led to the adoption of machine-learned interatomic potentials (MLIPs) as the MM models \cite{2021-qmmm3,grigorev2023calculation}. 

The concrete MLIPs ansatz we employ is the atomic cluster expansion (ACE) method~\cite{Drautz19, bachmayr19, 2020-pace1}, although we emphasise the workflow is readily transferable to other MLIPs approaches. The ACE model stands out for its ability to achieve high accuracy, despite being a {\it linear} model~\cite{2020-pace1}. The linear ACE model provides a parameterised site potential, for $\pmb{g}\in\big(\R^d\big)^{\Lhom\setminus 0}$, 
\begin{eqnarray}\label{eq:ACE_formula}
V^{\rm ACE}(\pmb{g}; \{c_B\}_{B\in\pmb{B}}) = \sum_{B\in\pmb{B}} c_B B(\pmb{g}), 
\end{eqnarray}
where $B$ are the basis functions and $c_B$ the parameters that we will estimate by minimizing a least square loss (cf.~\eqref{cost:forcemix}). The resulting ACE forces are denoted by $\F^{\rm ACE}_\ell(\pmb{g})$.

\subsubsection{The QM/MM hybrid model}

As detailed in \cite[Section 4]{2021-qmmm3}, the force-mixing approach is generally considered more computationally efficient and practically advantageous compared to the energy-mixing approach. We therefore focus on force-mixing schemes, while noting that our approach should be straightforward to adapt to energy-mixing methods. 

The approximated equilibrium state of QM/MM hybrid models is obtained by solving the following hybrid force balance equations: Find $\uH \in \Admu^{\rm H}(\L)$ such that
\begin{eqnarray}
\label{problem-f-mix}
	\F_{\ell}^{\rm H}\big(\bar{u}^{\rm H}\big) = 0
	\qquad \forall~\ell\in\LQM\cup\LMM,
\end{eqnarray}
where $\F_{\ell}^{\rm H}$ are the hybrid forces
\begin{eqnarray}
\label{F-H}
	\F_{\ell}^{\rm H}(u) =
	\left\{ \begin{array}{ll}
		\F^{\rm QM}_{\ell}(u), \quad & \ell\in\LQM
		\\[1ex]
		\F^{\rm ACE}_\ell(u),
		 \quad & \ell\in\LMM  
	\end{array} \right. ,
\end{eqnarray}
with $\F^{\rm QM}_{\ell}(u) := \F^{\LQM \cup \Lbuf}_{\ell}(u)$ and $\F^{\rm ACE}_\ell(u):=\F^{\rm ACE}\big( u_0(\cdot-\ell) + u(\cdot-\ell)\big)$.

To ensure the consistency of hybrid models (cf.~Remark~\ref{re:consis}), construct the ACE potential by employing the matching conditions derived and analyzed in~\cite{2021-qmmm3}. These matching conditions encode that the MM forces should match the QM forces to within a prescribed order off accuracy $K_{\rm F}$ in the limit of infinitesimal displacements. A rigorous justification of this approach is based on far-field regularity estimates on atomic displacement fields \cite{2021-qmmm3}.

Concretely, for a prescribed $K_{\rm F}\geq 1$, we estimate the ACE model parameters by minimising a least square loss that matches the QM to the ACE forces, as well as higher order derivatives, at the origin,
%
\begin{align}\label{cost:forcemix}
    \mathcal{L}_{\rm F}\big(\{c_B\}\big) &:= \sum_{j=0}^{K_{\rm F}} W^{\rm F}_j \big(\varepsilon_j^{\rm F}\big)^2 \\ 
    \notag 
    &:= \sum_{j=0}^{K_{\rm F}} W^{\rm F}_j \sum_{\pmb\ell = (\ell_1, ..., \ell_j)\in (\Lhom)^j}\Big|\F^{\rm h}_{,\pmb{\ell}}(\pmb 0) - \F^{\rm ACE}_{,\pmb{\ell}}(\pmb 0; \{c_B\})\Big|^2 {w}_j^{-1}(\pmb\ell).
\end{align}
Here, $w_j(\pmb\rho)$ is the weight function defined by  $w_j(\pmb \rho) := \prod_{i = 1}^j e^{-2\gamma|\rho_i|}$ with $\gamma$ a constant related to the locality of site potentials \assERL~(cf.~\cite[Section 3.1]{2021-qmmm3}) and $W_j^{\rm F}$ are additional empirical weights. The choices for $K_{\rm F}, \gamma$ and $W_j^{\rm F}$ will be specified in Section~\ref{sec:alg_numerics}.

It is worth noting that the choice of loss functional is not unique and several variants of \eqref{cost:forcemix} are available. For instance, one can incorporate the matching conditions for the virial stress of the QM reference model in the loss functional (see \eqref{cost:forcemix-dislocW}) to achieve improved accuracy. Further details can be found in \cite[Section 4]{2021-qmmm3}. The training sets and the hyperparameters for constructing the ACE basis will be tailored to the specific defective systems tested in Section~\ref{sec:alg_numerics}.

\begin{remark}\label{re:consis}
    A QM/MM model constructed as described above is {\it consistent} with the reference QM model in the following sense~\cite{2021-qmmm3}: Suppose that $\bar{u}$ is a strongly stable solution of~\eqref{eq:problem-force}, i.e., $\delta^2\E(\bar{u})$ is positive in $\UsH$. If the ACE models are constructed in a way that the $\ffit_j$ defined by~\eqref{cost:forcemix} are sufficiently small, then for sufficiently large QM and buffer regions, there exist equilibrium $\uH$ solving~\eqref{problem-f-mix}, such that
\begin{eqnarray}\label{ass:convergence_QMMM}
    \lim_{\NQM, \Nbuf \rightarrow \infty} \|\bar{u}-\uH\|_{\UsH} = 0.
\end{eqnarray}
    It is worth noting that the matching conditions between QM and MM models are crucial in constructing a consistent QM/MM scheme, which enables us to relate the following {\it a posteriori} residual estimates to {\it a priori} error estimates, as demonstrated in Lemma~\ref{lemma:res-F}.
\end{remark}

\section{A posteriori error estimates}
\label{sec:errest}

\subsection{An idealised error estimator}
\def\fR{\mathfrak{R}}
\def\etaR{\eta^{\fR}}
\def\etaS{\eta^{\mathfrak{S}}}
Following the analysis in \cite[Lemma 3.1]{CMAME} and adapting it to the force-mixing scheme using the techniques in \cite[Appendix C]{chen17}, we deduce that the {\it residual} force $\F(\uH)$ evaluated at a solution $\uH$ to \eqref{problem-f-mix} in dual norm characterizes the approximation error $\|\bar{u}-\uH\|_{\UsH}$. The proof can be found in the Appendix~\ref{sec:proofs}.


\begin{lemma}\label{lemma:res-F}
	Let $\bar{u}$ be a strongly stable solution of \eqref{eq:problem-force}. 
For $\varepsilon_1^{\rm F}$ sufficiently small and $\NQM, \Nbuf$ sufficiently large, there exists a solution $\uH$ of QM/MM force-mixing scheme~\eqref{problem-f-mix} and constants $c, C$ independent of the approximation parameters such that 
	\begin{eqnarray}\label{res-bound}
	c\|\bar{u}-\uH\|_{\UsH} \leq \| \F(\uH) \|_{(\UsH)^*} \leq C\|\bar{u}-\uH\|_{\UsH}.
	\end{eqnarray}
\end{lemma}

The estimate \eqref{res-bound} shows that the dual norm of the residual, $\| \F(\uH) \|_{(\UsH)^*}$, is a reliable and efficient {\it a posterior} error estimator. However, it is not directly computable and therefore not be used directly in practice.

In \cite[Section 3.3]{wang2020posteriori},  we solved a partial differential equation to evaluate $\| \cdot \|_{(\UsH)^*}$, which introduced an additional finite element mesh and led to significant additional algorithmic complexity. Here, we propose a more practical and more flexible approach, essentially replacing the PDE operator from \cite{wang2020posteriori} with a generalisation of a graph Laplacian. The operator we use was previously proposed in \cite{2016-precon1} for preconditioning geometry optimization. Algorithmically, this approach fits much better into the setting of atomistic modelling. In addition, the operator is ``aware'' of the atomic bonding and can more faithfully represent the discrete dual norm. 

To that end, we first construct the matrix $L^{\L}$ such that the elements satisfying
\begin{eqnarray}\label{LMatrix}
	L^{\L}_{ij} =
	\left\{ \begin{array}{ll}
		-\mu \exp\Big(-A\Big(\frac{|y_{ij}|}{y_{\rm nn}} - 1\Big)\Big), \quad & i \neq j, ~|y_{ij}| < r_{\rm cut}
		\\[0.3ex]
		0,  \quad & i \neq j,~|y_{ij}| \geq r_{\rm cut}
		\\[0.3ex]
		-\sum_{i\neq j} L_{ij} + \mu C_{\rm stab}, \quad & i = j
	\end{array} \right. ,
\end{eqnarray}
where $y_{ij} = y_i - y_j$ and the nearest-neighbour distance $r_{\rm nn}$ is obtained as the maximum of nearest neighbour bond lengths, i.e., $y_{\rm nn} = \max_{i}\min_{j\neq i} y_{ij}$. The parameters $\mu, A, r_{\rm cut}$ can be user-specified (e.g., $r_{\rm cut} = 2.5 y_{\rm nn}, A = 3$ appear to be a good defaults). The parameters can also be estimated numerically from the forces $\F_{\ell}$ to match the force jacobian as closely as possible.  
Given a specific connectivity defined by $r_{\rm cut}$ and setting $A = 0$ and $\mu = 1$, the matrix $L^{\L}$ simplifies to the standard graph Laplacian matrix. To enforce positive definiteness in practice, we stabilize $L^{\L}$ by introducing a diagonal term $C_{\rm stab}$, as suggested in \cite{2016-precon1}. The empirical selection of $C_{\rm stab} = 0.1$ has proven to outperform alternative choices in the numerical experiments. 

Let $\mathcal{F}(\uH)$ be the {\it residual} force evaluated at a solution $\uH$. We consider the equation
\begin{eqnarray}\label{eq:Lphi}
L^{\L} \cdot \phi_{\a}(\uH) = \mathcal{F}(\uH).
\end{eqnarray}
The following theorem establishes an equivalent representation of the residual force in dual norm through $\phi_{\a}(\uH)$, the solution to \eqref{eq:Lphi}. This result contributes to the derivation of the {\it ideal a posteriori} error estimator for the QM/MM scheme~\eqref{problem-f-mix}. We leave the proof to the Appendix~\ref{sec:proofs}.

\begin{theorem}\label{th:ctsphi}
	Let $\phi_{\rm a}$ be the solution to \eqref{eq:Lphi}. We define the {\it a posteriori} error estimator of the QM/MM scheme~\eqref{problem-f-mix} by
     \begin{equation}\label{eq:eta-ideal}
    	\eta^{\rm ideal}(\uH) := \|D \phi_{\a}(\uH)\|_{\ell^2_{\mathcal{N}}(\L)}.
    \end{equation}
    Then, there exist constants $c_1, C_1$ such that
	\begin{equation}\label{eq:equiv-1}
		c_1 \| \F(\uH) \|_{(\UsH)^*} \leq \eta^{\rm ideal}(\uH)
		\leq C_1 \| \F(\uH) \|_{(\UsH)^*}.
	\end{equation}
	Moreover, under the conditions of Lemma~\ref{lemma:res-F}, there exists another two constants $c_2, C_2$ such that 
	\begin{equation}\label{eq:equiv-2}
		c_2 \|\bar{u}-\uH\|_{\UsH} \leq \eta^{\rm ideal}(\uH)
		\leq C_2 \|\bar{u}-\uH\|_{\UsH}.
	\end{equation}
\end{theorem} 


In light with \eqref{eq:equiv-2}, the error estimator $\eta^{\rm ideal}(\uH)$ is referred to as {\it ideal} as it provides both upper and lower bounds for the approximation error. However, practical computation of $\eta^{\rm ideal}(\uH)$ remains challenging, primarily due to two main reasons. Firstly, the equation \eqref{eq:Lphi} is defined on an infinite lattice, posing difficulties for explicit solution. Secondly, obtaining the source term (residual force) $\F(\uH)$ is a non-trivial task.

\subsection{A practical error estimator}
\label{sec:sub:pra}

To overcome the aforementioned challenges, we propose the following two approximations: (1) truncating the infinite lattice $\L$ to a finite lattice; (2) constructing the approximated force $\widetilde{\F}(\uH)$ through a linear expansion of the residual force $\F(\uH)$ away from the defect core.

\subsubsection{Truncation}
\label{sec:sub:trun}
\def\RO{R_{\Omega}}
\def\T{\mathcal{T}}
Let $\Omega$ be a convex polygon or polyhedron in $\R^d$ with boundary $\partial\Omega$ such that $\L^{\rm QM}\cup \L^{\rm MM} \subset \Omega$ and $\L^{\Omega}:=\Lambda\cap\Omega$. We interpret the lattice $\Lambda$ as the vertex set of a simplicial grid $\T$ leading to the {\it canonical partition}. For simplicity, 
suppose that $\Omega$ is compatible with $\T$, i.e., there exists a subset $\T_{\Omega} \subset \T$ such that $\text{clos}(\Omega)=\cup \T_{\Omega}$.
As a truncation of \eqref{LMatrix}, we denote the corresponding truncated Laplace matrix as $L$.

The error resulting from this truncation has already been analyzed in \cite[Section 3.5]{wang2020posteriori} and in principle one can incorporate this to adapt the size of MM region. However, the adaptation of QM and buffer regions is much more important in practice. Therefore, in this paper, we assume that the computational domain $\Omega$ (or equivalently $\LMM$) is chosen to be sufficiently large to ensure that the corresponding truncation error is negligible for the error estimator.

\subsubsection{Approximated residual force}
\label{sec:sub:f}

To reduce the computational cost of evaluating the residual force $\F(\uH)$ while maintaining its accuracy, an approximation scheme should be carefully designed. We note that the exact evaluation of residual force is necessary for accurate adjustment of the QM and buffer regions, which is in fact acceptable since the size of $\L^{\rm QB}:= \Lambda^{\rm QM }\cup \Lbuf$ is much smaller than that of the full computational domain $\Omega$. We therefore only need to consider the approximation of the residual force outside this region, where the displacement $\uH$ varies smoothly.

To that end, we consider the linear approximation of residual force. Let $\delta \F^{B_{R_{\rm buf}}}({\bm 0})$ be the QM force constant defined on the homogeneous lattice inside the region $B_{R_{\rm buf}}$.
The residual force is then approximated in a mixed scheme
\begin{align}\label{eq:ACE2F}
	\widetilde{\F}_{\ell}(\bar{u}^{\rm H}) =
	\left\{ \begin{array}{ll}
		\F_{\ell}(\bar{u}^{\rm H}) \qquad & \ell\in \L^{\rm QB} 
		\\[1ex]
		\delta \F^{B_{R_{\rm buf}}}({\bm 0}) \Big(\big(u_0(\cdot-\ell) + \uH(\cdot - \ell)\big)\big|_{B_{R_{\rm buf}}(\ell)}\Big) \qquad 
		 & \ell\in \Lambda^{\Omega} \setminus \L^{\rm QB}
	\end{array} \right. . 
\end{align}
Due to the locality of QM site potentials (cf. \assERL~and \cite[Lemma 2.1]{chen17}), the accuracy of this approximation can therefore be guaranteed as long as the size of core region $\L^{\rm QB}$ and $\Rbuf$ are chosen to be sufficiently large. The error caused by this approximation can be analyzed following the approach in \cite[Section 5.1]{wang2020posteriori}, referred to as the data oscillation term. Although we omit the details here, it is important to note that this error can be controlled.

Hence, combined with the truncation and the approximation of the residual force, we obtain the approximation to \eqref{eq:Lphi}, which is expressed as
\begin{eqnarray}
\label{eq:stress_force_trun_T}
 L \cdot \phi(\uH) = \widetilde{\F}(\uH).
\end{eqnarray}
The {\it practical a posteriori} error estimator of the QM/MM scheme~\eqref{problem-f-mix} is then given by
\begin{equation}\label{eq:eta}
\eta(\uH) := \|D \phi(\uH)\|_{\ell^2_{\mathcal{N}}(\L^{\Omega})}.
\end{equation}
We will utilize it to formulate the main adaptive QM/MM algorithm in the following section.

In practical implementations, it is essential to accurately compute the residual force $\mathcal{F}_{\ell}(\uH)$ within $\Lambda^{\rm QB}$. To achieve this, an extended buffer region, twice the size of the original buffer region $R_{\rm buf}$, is employed for computing the residual force $\mathcal{F}_{\ell}(\uH)$. This practice becomes particularly crucial when employing the electronic structure model as the reference model (cf.~Section~\ref{sec:sub:sub:dft}).

\section{Adaptive algorithm and numerical tests}
\label{sec:alg_numerics}

In this section, we present an adaptive QM/MM algorithm leveraging the {\it practical a posterior} error estimator $\eta(\uH)$ defined by \eqref{eq:eta}, and conduct the numerical examples for three typical crystalline defects: in-plane crack in W, (110)[100] edge dislocation and di-interstitial in Si. 

\subsection{Adaptive algorithm}
\label{sec:sub:alg}

The basic idea of our adaptive QM/MM algorithm is to repeat the following procedure before reaching the required accuracy:
$$
\mbox{Solve}~\rightarrow~\mbox{Estimate}~\rightarrow~
\mbox{Mark \& Refine}.
$$
Compared to standard adaptive finite element methods~\cite{Dorfler:1996}, our approach differs in that we integrate the ``Mark" and ``Refine" steps by employing a fast marching method for updating the QM and buffer regions, as presented in Algorithm \ref{alg:fmm}. Moreover, as discussed in Section \ref{sec:sub:trun}, the computational domain $\Omega$ (or equivalently, MM region $\LMM$) is fixed to be sufficiently large to mitigate the influence of the truncation error. Hence, in the ``Mark \& Refine" step, we only update QM and buffer regions.

To choose where to refine the models, it is natural to assign the global estimator $\eta(\uH)$ into local contributions (site-based),
\begin{eqnarray}\label{eq:local_eta}
\eta_{\ell}(\uH):= \frac{\big|D\phi(\uH)(\ell)\big|^2_{\mathcal{N}}}{\eta(\uH)} \qquad \forall \ell \in \L^{\Omega},
\end{eqnarray}
where $|\cdot|_{\mathcal{N}}$ is defined by \eqref{eq: nn norm} and it is straightforward to see that $\sum_{\ell \in \L^{\Omega}} \eta_{\ell}(\uH)=\eta(\uH)$. 

We first describe the adaptive QM/MM algorithm as follows and then give a detailed discussion. 

\begin{algorithm}[H]
\caption{Adaptive QM/MM algorithm}
\label{alg:main}
\hskip-6cm {\bf Prescribe} $\LQM, \Lbuf, \LMM$, termination tolerance $\eta_{\rm tol}$.
\begin{algorithmic}[1]
\Repeat
	\State{ \textit{Solve}: Solve \eqref{problem-f-mix} to obtain $\uH$. }
	\State{ \textit{Estimate}: Compute $\eta(\uH)$ and $\eta_{\ell}(\uH)$ by \eqref{eq:eta} and \eqref{eq:local_eta} respectively. } 		
	\State{ \textit{Mark \& Refine}: Apply Algorithm \ref{alg:fmm} to construct new $\LQM$ and $\Lbuf$ regions.}
\Until{$\eta(\uH) < \eta_{\rm tol}$}
\end{algorithmic}
\end{algorithm}

Given a partition $\LQM$, $\Lbuf$ and $\LMM$, the ``Solve" step computes the approximated equilibrium state $\uH$ by solving~\eqref{problem-f-mix}.
The ``Estimate" step evaluates the {\it practical a posteriori} error estimator~\eqref{eq:eta} and its local contribution~\eqref{eq:local_eta}. The ``Mark \& Refine" step is fundamental in developing a robust adaptive algorithm, warranting an in-depth discussion, as outlined below.

{\bf Mark \& Refine.} 
In place of the isotropic approach detailed in~\cite[Section 4.1]{wang2020posteriori}, where the radii of QM and buffer regions are adjusted, we introduce a more robust algorithm specifically designed for anisotropic defects. This method entails solving an interface motion problem (cf.~\eqref{eq:eikonal}) using the fast marching method~\cite{chopp2001some, sethian1999fast} to manage the evolution of QM/MM partitions, all while incorporating the {\it a posteriori} error estimator introduced in the last section. 

The fast marching method is a numerical technique for finding the solutions of the Eikonal equation~\cite{zhao2005fast},
\begin{equation}\label{eq:eikonal}
    G(x) | \nabla T(x) | = 1,
\end{equation}
where $T(x)$ is the first arriving time of a {\it closed} interface at $x\in\R^d$ and $G(x)>0$ is the speed in the normal direction at $x$.

Adapting this method into our adaptive QM/MM scheme involves several key steps. First, we define the interface of QM region. Next, the local error estimator~\eqref{eq:local_eta} needs to be transformed into the speed function $G$. Finally, we consider the geometric relationship between lattice sites and the meshgrid to numerically apply the fast marching method. These three aspects will be addressed in the following algorithm, which implements the ``Mark \& Refine" step and completes the specification of our adaptive QM/MM algorithm (Algorithm~\ref{alg:main}).

\begin{algorithm}[H]
\caption{Fast marching method for marking and refining}
\label{alg:fmm}
\hskip-5.75cm {\bf Prescribe} $\Omega_{\rm est}$, $C_{\rm new}$, $C_{\rm buf}$, $\LQM$, and local error estimator $\eta_{\ell}(\uH)$.
\begin{algorithmic}[1]
	\State{ \textit{Construct meshgrid}: Generate an equal spaced meshgrid $\T_{\rm est}$ in $\Omega_{\rm est}$.}
	\State{ \textit{Define QM interface}: Construct a scalar function $\varphi$ defined on $\T_{\rm est}$ such that the QM {\it interface} is defined as the zero contour of $\varphi$.}
	\State{ \textit{Construct speed}: Establish the discrete speed function $G$ based on the scattered interpolation of the local error estimator $\eta_{\ell}(\uH)$ from lattice sites to the meshgrid $\T_{\rm est}$.}
    \State{ \textit{Solve arriving time}: Solve \eqref{eq:eikonal} to obtain the first arriving time $T$ on $\T_{\rm est}$. Interpolate it back to lattice sites.}
    \State{ \textit{Determine new regions}: Update $\LQM_{\rm new}$ and $\Lbuf_{\rm new}$ by \eqref{eq:determine_new_rgn} with two thresholds $C_{\rm new}$ and $C_{\rm buf}$.}
\end{algorithmic}
\hskip-13.2cm {\bf Output:} $\LQM_{\rm new}, \Lbuf_{\rm new}$.
\end{algorithm}

We give a detailed discussion of the individual steps in Algorithm~\ref{alg:fmm}.

{\it Construct meshgrid:} To employ the fast marching method, we first construct an equi-spaced meshgrid $\T_{\rm est}$ in a cubic (square for $d=2$) region $\Omega_{\rm est}$ satisfying $\L^{\rm QB} \subset \Omega_{\rm est}$. Since we only update the QM and buffer regions, the Eikonal equation \eqref{eq:eikonal} is therefore only solved in $\Omega_{\rm est}$ to avoid unnecessary computational cost. Let $\mathcal{N}_{\rm est}$ be the set of the nodes of $\T_{\rm est}$ and $h_{\rm est}$ be the length of mesh. 

{\it Define QM interface:} We define the QM interface based on $\LQM$. A scalar function $\varphi$ defined on $\mathcal{N}_{\rm est}$ is constructed such that the current QM interface is a zero contour of $\varphi$. In particular, we first find the geometric relationship between $\mathcal{N}_{\rm est}$ and $\LQM$, and then let $\varphi=-1$ when the nodes in $\mathcal{N}_{\rm est}$ belong to the inside QM region, otherwise $\varphi=1$. As a result, the zero contour of $\varphi$ can be generated automatically (cf. Figure~\ref{fig:refine}). 

{\it Construct speed:} We construct the speed function $G$ discretized by $\T_{\rm est}$, based on the scattered interpolation of the local error estimator $\eta_{\ell}(\uH)$ from the lattice sites $\L^{\rm QB}$ to the meshgrid nodes $\mathcal{N}_{\rm est}$ (cf. Figure~\ref{fig:apevsspeed}). To be more precise, for $\ell \in \L^{\rm QB}$, given a nodal basis function $\Phi_{\ell}$, for any $n_{\rm est} \in \mathcal{N}_{\rm est}$, we define
\begin{eqnarray}\label{eq:scattered_interp}
G(n_{\rm est}) := \sum_{\ell \in \L^{\rm QB}} \eta_{\ell}(\uH)\Phi_{\ell}(n_{\rm est}).
\end{eqnarray}
We utilize linear basis function throughout our numerical experiments. 

{\it Solve arriving time:} Then, we apply the fast marching method to obtain the first arriving time $T$ by solving the Eikonal equation \eqref{eq:eikonal} on $\mathcal{N}_{\rm est}$. In practice, we use a python extension module named {\tt scikit-fmm} \cite{gitfmm} to realize the fast marching method. As $T$ is defined on $\mathcal{N}_{\rm est}$, again, the scattered interpolation is utilized to evaluate the first arriving time at each atom $\ell \in \L\cap \Omega_{\rm est}$. 

{\it Determine new regions:} Given two thresholds $T_{\rm new}$ and $T
_{\rm buf}$, the new QM and buffer regions $\L^{\rm QM}_{\rm new}$ and $\L^{\rm BUF}_{\rm new}$ are then determined (cf.~Figure~\ref{fig:refine}) such that 
\begin{equation}\label{eq:determine_new_rgn}
    T(\L^{\rm QM}_{\rm new}) \leq T_{\rm new}, \qquad
    T_{\rm new} < T(\L^{\rm BUF}_{\rm new}) \leq T_{\rm new}+T_{\rm buf}, 
\end{equation}
where $T(\L_{*}):=\max_{\ell\in\L_{*}}\{T(\ell)\}$ with  $\L_{*}=\L^{\rm QM}_{\rm new}, \L^{\rm BUF}_{\rm new}$. These two thresholds will be studied in the next section.

We summarize that all simulations presented in this work are implemented in several open-source {\tt Julia} packages: {\tt SKTB.jl} \cite{gitSKTB} (for the NRL tight binding model), {\tt ACEpotentials.jl} \cite{witt2023acepotentials, gitACEpotentials} (for the construction of ACE basis and the fitting of ACE models), {\tt QMMM2.jl} \cite{gitQMMM2} (for the QM/MM coupling scheme) and {\tt AdapQMMM.jl} \cite{gitadapQMMM} (for the adaptivity). All tests we report except the DFT simulation (cf.~Section \ref{sec:sub:sub:dft}) are performed on an {\tt Intel(R) Core(TM) i7-7820HQ CPU @2.90GHz}, with {\tt macOS (x86-64-apple-darwin19.6.0)} operating system. The DFT simulation is simulated on a Linux cluster with {\tt AMD EPYC-Rome Processor} with 96 cores and 1TB memory. 

\subsection{Adaptive algorithm study}
\label{sec:sub:params_study}

In this section, we conduct a detailed study of the performance of our adaptive algorithm for (001)[100] edge dislocation in W. The illustration of domain decomposition for this case has already shown in Figure~\ref{fig:decomposition}. In order to test our adaptive algorithms in the simplest possible setting we use an embedded atom model (EAM) \cite{Daw1984a} as the reference model instead of an actual QM model. This allows us to explore the algorithms in a wider parameter range.

In light of the theory in~\cite{2021-qmmm3}, for dislocation simulations, the training set of constructing the ACE potential for {\it consistent} QM/MM methods (cf.~\eqref{eq:Wdisloc}) should contain the first-order derivative of the QM force (QM force constant $\delta f({\bf 0})$) as well as the second-order derivative of the virial ($\partial^3_{\mathsf{F}}W^{\rm h}_{\rm cb}(\mathsf{I})$) evaluated on the homogeneous lattice $\Lhom$. The loss function~\eqref{cost:forcemix} is then given by
\begin{align}\label{cost:forcemix-dislocW}
\mathcal{L}_{\rm FV}\big(\{c_B\}\big) := W^{\rm F}_1 \big(\varepsilon_1^{\rm F}\big)^2 + W^{\rm V}_2 \big(\varepsilon_2^{\rm V} \big)^2
\end{align}
with $\varepsilon_1^{\rm F}$ defined by \eqref{cost:forcemix} and 
\[
\vfit_2 := \big|\partial^{3}_{\mathsf{F}} \Wcb^{\rm h}(\mathsf{I}) - \partial^{3}_{\mathsf{F}} \Wcb^{\rm ACE}(\mathsf{I}) \big| :=\Bigg| \sum_{\pmb\rho \in (\Lhom\setminus 0)^{3}} \big(V^{\rm h}_{,\pmb\rho}(\mathsf{I}) - V^{\rm ACE}_{,\pmb\rho}(\mathsf{I})\big) \otimes {\pmb\rho} \Bigg|
\]
where $\Wcb^{\rm h}$ and $\Wcb^{\rm ACE}$ are the corresponding Cauchy-Born elastic energy density functional~\cite[Eq.(3.4)]{2021-qmmm3} and $\mathsf{I} \in \R^{d\times d}$ is the identity matrix. The weights $W^{\rm F}_1=10$ and $W^{\rm V}_2=1$ are originally taken from \cite[Table F.1]{2021-qmmm3}.

It is shown in \cite[Theorem 3.4]{2021-qmmm3} that the {\it a priori} error estimate for the corresponding QM/MM force-mixing for edge dislocations gives
\begin{eqnarray}\label{eq:Wdisloc}
\|\bar{u} - \uH\|_{\UsH} \lesssim N_{\rm QM}^{-1}.
\end{eqnarray}

The loss function \eqref{cost:forcemix-dislocW} is quadratic in the parameters $\{c_B\}$ and can therefore be minimised using Bayesian linear regression schemes. In our implementation we employ the Automatic Relevance Determination (ARD) \cite{wipf2007new} to achieve the parameter estimation, which is a known statistical technique used to automatically determine the relevance of input features or variables in a predictive model.

\begin{figure}[!htb]
        \centering
        \subfloat[$\tilde{\eta}_{\ell}(\uH)$ \label{fig:ape}]{
		\includegraphics[height=6.0cm]{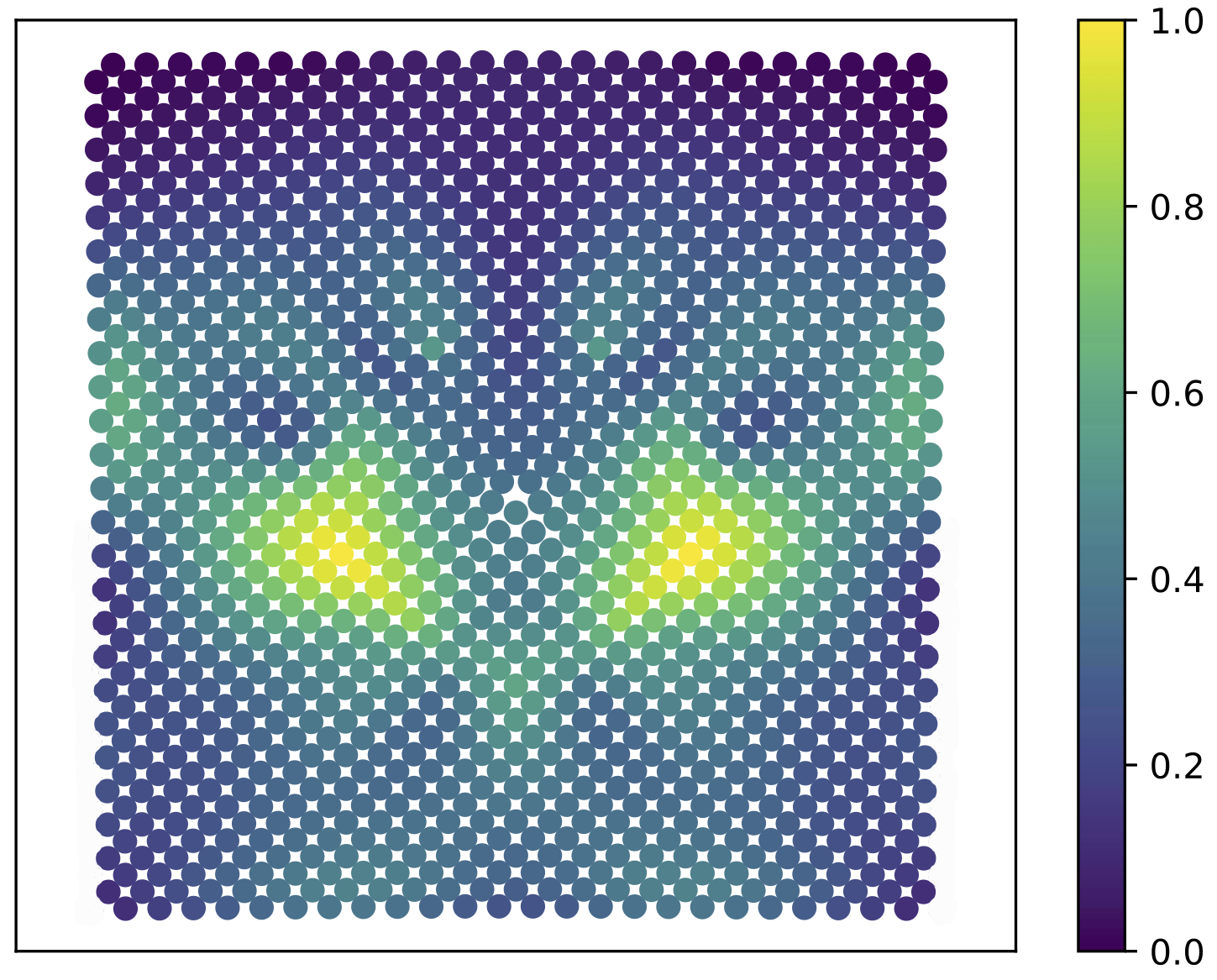}}
		\quad
        \subfloat[$G(n_{\rm est})$ \label{fig:speed}]{
        \includegraphics[height=6.0cm]{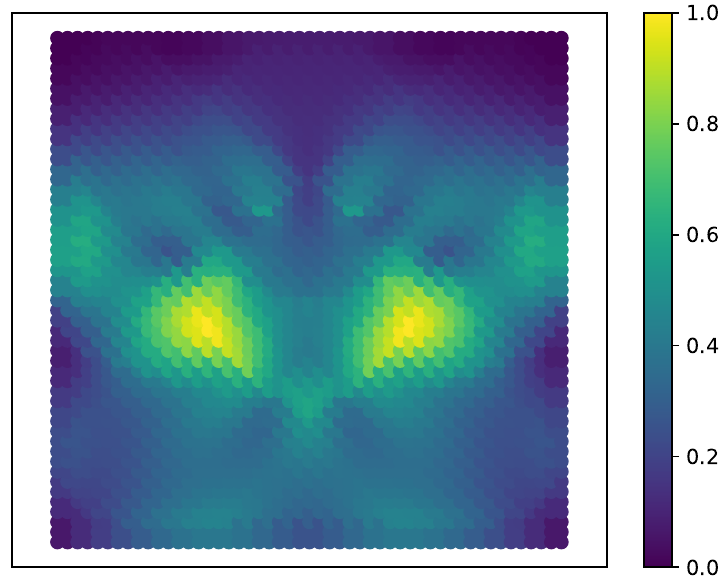}}
        \caption{The distributions of the {\it a posteriori} error estimator (left) defined on $\L_{\rm est}$ and the speed (right) defined on $\mathcal{N}_{\rm est}$.}
        \label{fig:apevsspeed}
\end{figure}

{\it Algorithm \ref{alg:fmm}:} Figure \ref{fig:apevsspeed} visualises the ``Construct speed" step in Algorithm \ref{alg:fmm}. The error estimator $\eta_{\ell}(\uH)$ defined on the lattice sites in $\L_{\rm est}$ are shown in the left figure while the corresponding speed $G(n_{\rm est})$ constructed by the scattered interpolation of $\tilde{\eta}_{\ell}(\uH)$ (i.e., the normalization of $\eta_{\ell}(\uH)$) from $\L_{\rm est}$ to $\mathcal{N}_{\rm est}$ (cf.~\eqref{eq:scattered_interp}) are presented as the right panel. As we can see, the scattered interpolation behaves well if $h_{\rm est}=0.4$\AA~is chosen to be small enough compared to the lattice constant of given species ($r^{\rm W}_{0}=2.73$\AA). 

\begin{figure}[!htb]
        \centering
        \includegraphics[height=6cm]{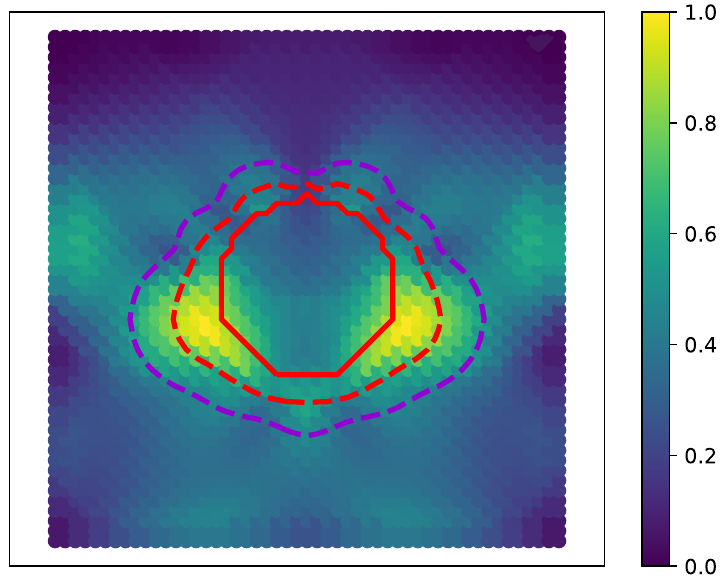}
        \caption{The illustration of the ``Mark \& Refine" step plotted on $\mathcal{N}_{\rm est}$: current QM interface (red solid line), new QM interface (red dashed line) and new buffer interface (purple dashed line). Colors represent the normalization of error estimator (speed) defined on $\mathcal{N}_{\rm est}$.}
        \label{fig:refine}
\end{figure}

Figure \ref{fig:refine} illustrates the ``Determine new regions" step in Algorithm \ref{alg:fmm}, drawing the interface of QM and buffer regions defined on the meshgrid $\mathcal{N}_{\rm est}$. The current QM interface (red solid line) is constructed by the ``Define QM interface" step in Algorithm \ref{alg:fmm}. The speed $G$, determined via ``construct speed" step, is visualized by color. Solving the corresponding Eikonal equation \eqref{eq:eikonal} using the fast marching method with parameters $h_{\rm est}=0.4$\AA~and $T_{\rm new}=T_{\rm buf}=3.0$,
the new QM interface (red dashed line) and the new buffer interface (purple dashed line) determined by the strategy \eqref{eq:determine_new_rgn} are also presented in this figure. We observe that our algorithm automatically adjusts the QM/MM partitioning anisotropically based on the error estimator. The specific choice of the those parameters will be discussed next. 

{\it The mesh size $h_{\rm est}$ and the thresholds $T_{\rm new}, T_{\rm buf}$:} The parameters involved in Algorithm \ref{alg:fmm} have a significant impact on its performance, and then in turn affect the overall performance of our {\em outer} adaptive algorithm (Algorithm \ref{alg:main}). It is therefore essential to carefully choose and control these parameters to ensure a robust and efficient implementation of the algorithm. More specifically, the choice of $h_{\rm est}$ mainly influences the accuracy of the scattered interpolation for constructing the speed $G$. In this study, we set $h_{\rm est}=0.2$\AA~to ensure that it is sufficiently small relative to the lattice constant. This choice minimizes the interpolation error compared to the {\it a posteriori} error estimator. The effect of $T_{\rm buf}$ on the adaptive computations is marginal due to the locality of QM site potentials~\assERL. Hence, we fix $T_{\rm buf}=3.0$~in Algorithm \ref{alg:fmm} throughout the numerical experiments. Next, we study the sensitivity of the main algorithm (Algorithm \ref{alg:main}) to the parameter $T_{\rm new}$ shown in Algorithm \ref{alg:fmm}. 

To that end, since we are now using the EAM potential as the reference model, the residual forces $\F(\uH)$ can be evaluated exactly. Hence, to mitigate the impact of errors from other sources, we can directly obtain the {\it exact} error estimator by solving \eqref{eq:stress_force_trun_T} with the exact residual forces instead of the approximated residual forces $\widetilde{\F}(\uH)$ introduced in Section \ref{sec:sub:f}, namely
\begin{eqnarray}\label{eq:exact}
L \cdot \phi_{\a}^{\rm exact} = \F(\uH)
\end{eqnarray}
Denote $\eta^{\rm exact}(\uH):=\|D \phi^{\rm exact}_{\rm a}\|_{L^2(\L^{\Omega})}$. The remaining error between $\|D \phi^{\rm exact}_{\rm a}\|_{\ell^2_{\mathcal{N}}(\L^{\Omega})}$ and the idealised estimator $\|D\phi\|_{L^2(\L)}$ is the domain truncation error, which is sufficiently small when $\Omega$ is fixed to be large (cf. \cite[Section 4.3]{wang2020posteriori}). Moreover, compared with \eqref{eq:stress_force_trun_T} and \eqref{eq:exact}, the difference between $\eta(\uH)$ and $\eta^{\rm exact}(\uH)$ in fact indicates the accuracy of the approximation of residual force, which will be numerically verified in the next section.

\begin{figure}[!htb]
    \centering
    \subfloat[$T_{\rm new}=1.0$\label{fig:study_C_1}]{
    \includegraphics[height=4.2cm]{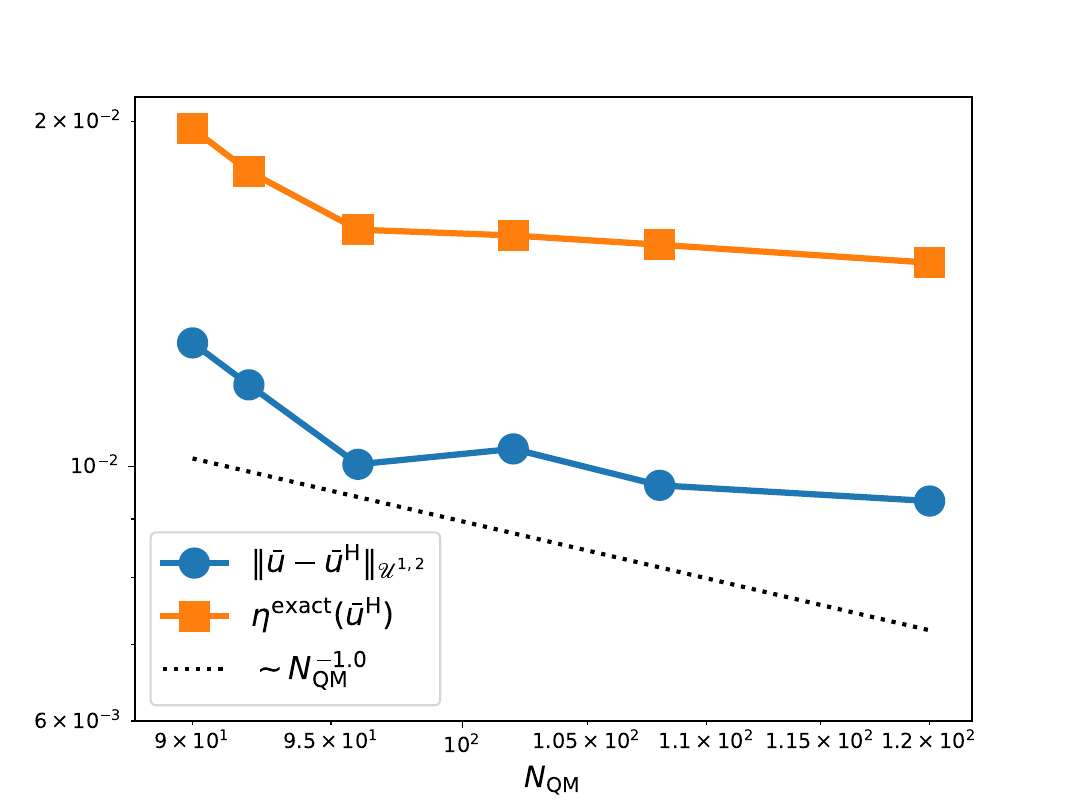}} 
    \subfloat[$T_{\rm new}=4.0$ \label{fig:study_C_2}]{
    \includegraphics[height=4.2cm]{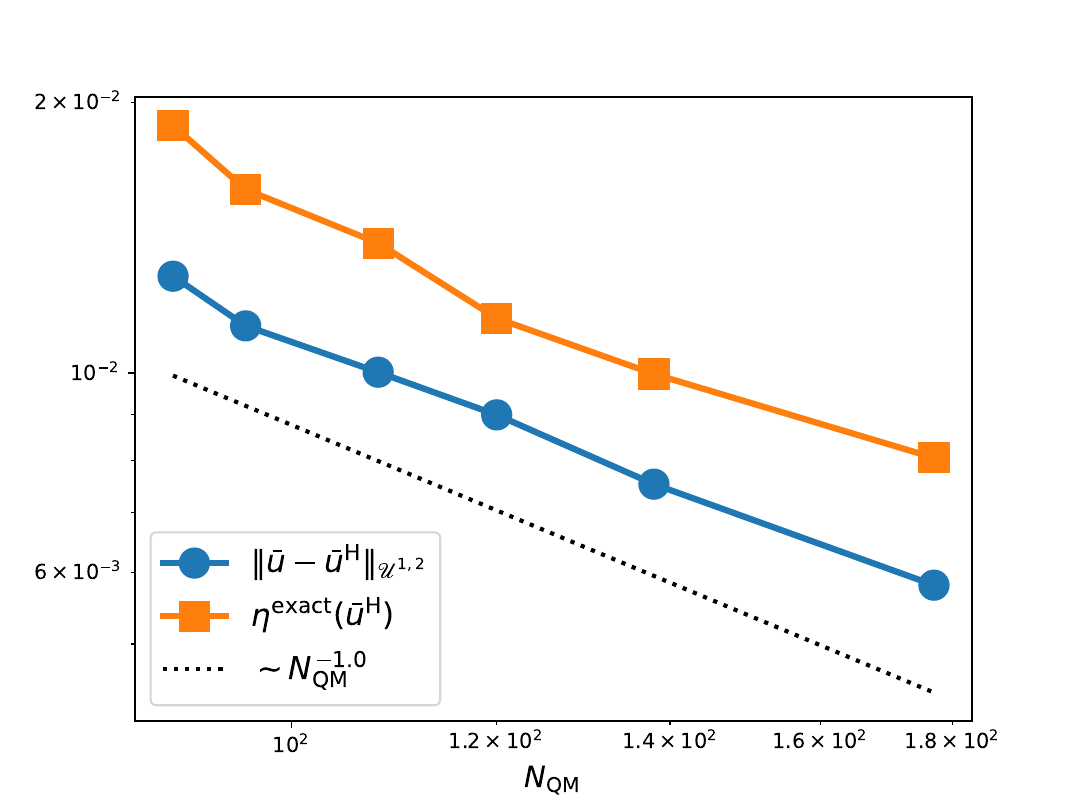}} 
    \subfloat[$T_{\rm new}=7.0$ \label{fig:study_C_3}]{
    \includegraphics[height=4.2cm]{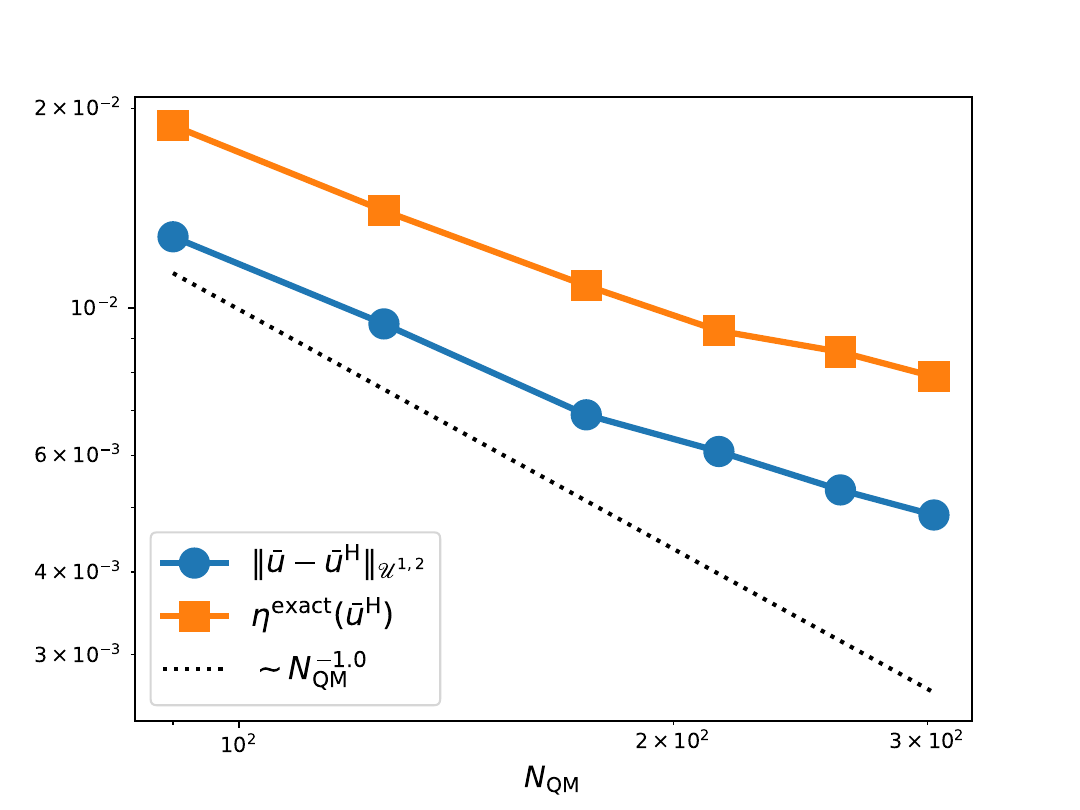}} 
    \caption{QM/MM errors and error estimators plotted against $N_{\rm QM}$ in the adaptive QM/MM Algorithm \ref{alg:main}. Different $T_{\rm new}$ employed in Algorithm~\ref{alg:fmm} can lead to qualitatively different behaviour in the adaptive computations.}
    \label{fig:study_C}
\end{figure}

Figure \ref{fig:study_C} shows that the convergence of Algorithm \ref{alg:main} for (001)[100] edge dislocation in W with different $T_{\rm new}$. In each figure we plot the approximation error $\|\bar{u}-\uH\|_{\UsH}$ and the {\it exact} error estimator $\eta^{\rm exact}(\uH)$ against $N_{\rm QM}$. It demonstrates that different threshold $T_{\rm new}$ employed in Algorithm \ref{alg:fmm} can lead to qualitatively different behaviour in the QM/MM model refinement. From the strategy \eqref{eq:determine_new_rgn}, the value of $T_{\rm new}$ reflects the increment of current QM region based on the error estimator. Both large or small $T_{\rm new}$ will lead to the sub-optimal convergence for the reason that the QM/MM subsystems are not optimally assigned. Figure \ref{fig:study_C} numerically demonstrates that $T_{\rm new}=4.0$~gives the optimal rate of convergence and the best agreement between the approximation error and the error estimator. Note that the sensitivity of $T_{\rm new}$ to the choice of system is not significant. Hence, we fix $T_{\rm new}=4.0$~for all benchmark problems in the remaining part of this paper. %

\subsection{In-plane crack in W}
\label{sec:sub:numerics_crack}

\begin{figure}[!htb]
    \centering
    \includegraphics[height=5cm]{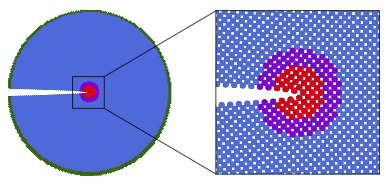}
    \caption{Domain decomposition for in-plane crack in W.}
    \label{fig:geom_crack}
\end{figure}

In this study, we investigate the behavior of our adaptive QM/MM algorithm (Algorithm \ref{alg:main}) for an in-plane crack in W. In order to test our adaptive algorithms in the simplest possible setting we use an embedded atom model (EAM) \cite{Daw1984a} as the reference model instead of an actual QM model. This allows us to explore the algorithms in a larger computational domain and still compare against a solution with the exact reference model. The far-field predictor, $u_0$, for the crack is briefly formulated in Appendix \ref{sec:sub:apd:crack}. Note that the use of an EAM potential as the reference model, as opposed to an actual ab initio model, enables us to perform large-scale tests more easily. 
We apply a quasi-2D setting, where clamped boundary conditions are used in the (001) plane and periodic boundary conditions in the [001] direction. The illustration of the QM/MM decomposition and core geometry is given in Figure \ref{fig:geom_crack}. Furthermore, we choose the radii of the computational domain $\Omega$ and the MM region $\LMM$ to be sufficiently large, specially $R_{\Omega}=120r^{\rm W}_0$ and $\RMM=110r^{\rm W}_0$, where $r^{\rm W}_0$ denotes the lattice constant of W (BCC). 

\begin{figure}[!htb]
    \centering
    \includegraphics[height=6cm]{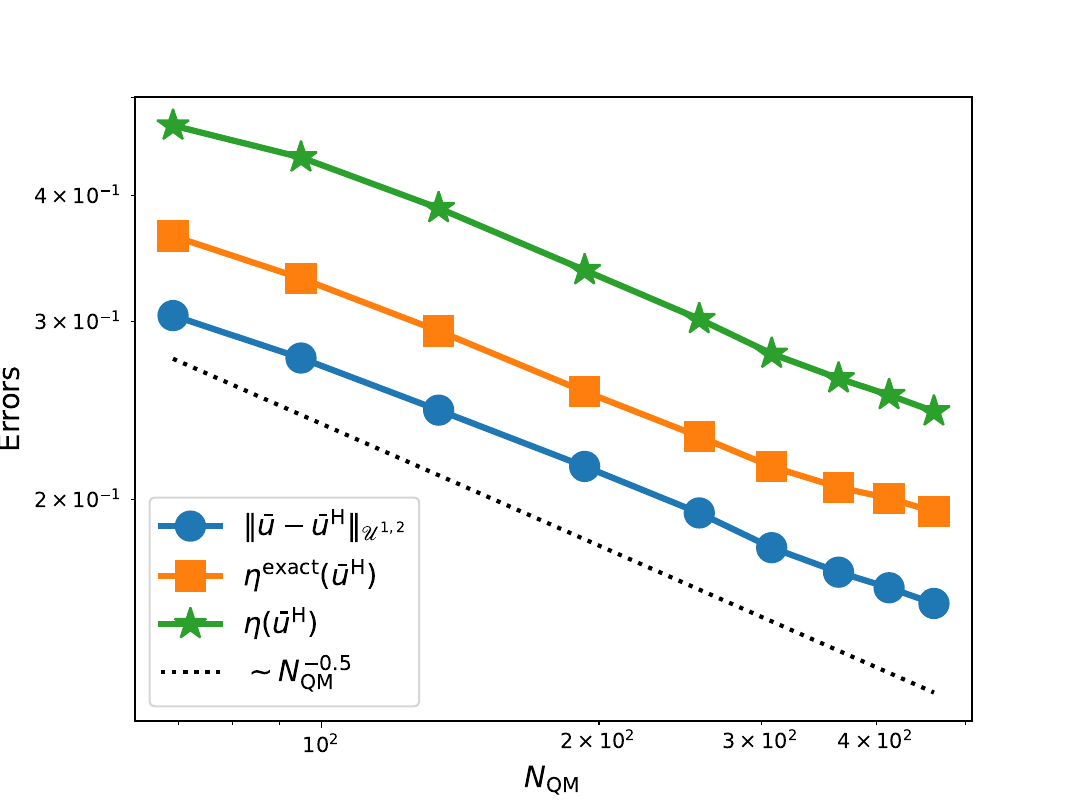}
    \caption{EAM W model: Convergence of the adaptive Algorithm \ref{alg:main} for in-plane crack.}
    \label{fig:conv_crack}
\end{figure}

In order to construct the ACE potential for QM/MM models for a crack, we follow the same approach as that for dislocation introduced in Section \ref{sec:sub:params_study}. The training set for the ACE potential contains the same observations, i.e., the QM force constant and the second order derivatives of the virial evaluated on the homogeneous lattice. Furthermore, random surface configurations are incorporated into the training set along with total energies and forces, with a weight ratio of 10:1, to account for surface effects.  While a rigorous {\it a priori} error estimate for QM/MM models for crack is still lacking, it is reasonable to speculate that an estimate analogous to the dislocations case given in~\eqref{eq:Wdisloc},  holds, 
\begin{eqnarray}\label{eq:apriori_crack}
\|\bar{u} - \uH\|_{\UsH} \lesssim N_{\rm QM}^{-0.5},
\end{eqnarray}
where we exploit the fact that the decay of the far-field predictor for crack is $|D u_0(\ell)|\lesssim |\ell|^{-0.5}$ (while it was $|D u_0(\ell)|\lesssim |\ell|^{-1}$ for a straight dislocation). The sketch of the proof is briefly given in the Appendix~\ref{sec:proofs} while the numerical verification is shown in Figure~\ref{fig:conv_crack}. 

We then study the convergence of Algorithm~\ref{alg:main} for an in-plane crack in W. In Figure~\ref{fig:conv_crack} we plot the approximation error $\|\bar{u}-\uH\|_{\UsH}$, and the error estimators $\eta(\uH)$, $\eta^{\rm exact}(\uH)$ against $N_{\rm QM}$. We observe two things: First, the two error estimators follow the trend of the approximation
error fairly closely, which confirms that the practical estimator $\eta(\uH)$ and the (nearly) ideal estimator $\eta^{\rm exact}(\uH)$ provide efficient and reliable estimators for the QM/MM model residual. Secondly, the difference between $\eta(\uH)$ and $\eta^{\rm exact}(\uH)$ is marginal (with a prefactor 1.32), confirming the accuracy of the approximation of residual force introduced in Section~\ref{sec:sub:f}.

In this numerical case, the CPU time in seconds required for simulating different components (steps) in Algorithm~\ref{alg:main} is not provided. As the empirical EAM potential serves as the reference model, the cost of computing the exact residual forces is relatively inexpensive. However, when using an actual QM model (such as DFT or NRLTB~\cite{cohen94}), the cost of evaluating the exact residual forces will significantly increase compared to computing the QM/MM solution $\uH$. This will be clearly illustrated in the following numerical examples (refer to Figure~\ref{fig:time_disloc_Si} and Figure~\ref{fig:time_diint_Si}).



\begin{figure}[!htb]
    \centering
    \includegraphics[height=3.75cm]{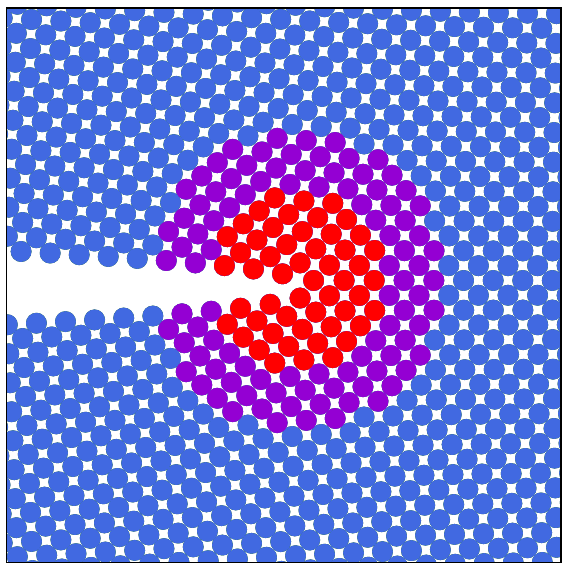}
    \includegraphics[height=3.75cm]{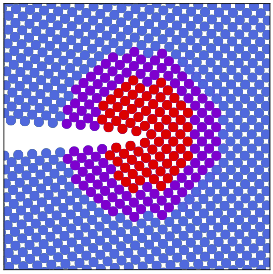}
    \includegraphics[height=3.75cm]{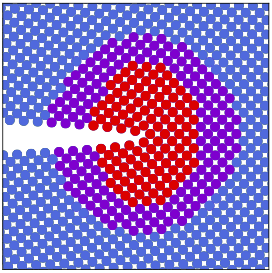}
    \includegraphics[height=3.75cm]{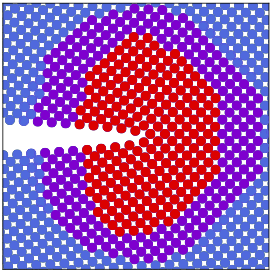}
    \caption{Evolution of QM and MM partitions in the adaptive procedure for in-plane crack in W. $N_{\rm QM}$ are 56, 91, 122, 183 from left to right.}
    \label{fig:evolution_crack}
\end{figure}

Figure~\ref{fig:evolution_crack} presents the evolution of the QM/MM partitions throughout the adaptation process. The initial geometry of the system features an isotropic QM region. As the adaptive computations progress, the QM region is anisotropically adjusted in accordance with the {\it a posteriori} error estimator. This demonstrates the robustness and adaptability of our main algorithm (Algorithm~\ref{alg:main}).

\subsection{Edge dislocation in Si}
\label{sec:sub:numerics_edge}

Next, we test our adaptive QM/MM schemes when the QM reference model is a simple electronic structure model for Si. We choose the NRL-TB model \cite{cohen94} as the reference model, which is a successful tight-binding model.  See the Appendix~\ref{sec:NRL} for a short review. Its much lower computational cost (compared with DFT) allows us to perform some validation that would no longer be possible with DFT. Our choice of Si as the material is due to the fact that it is a rich semi-conducting material for which we have also strong theoretical and numerical evidence for the localisation of its interatomic forces~\cite{chen18, chen16}.

We consider a (110)[100] edge dislocation in Si, where the same quasi-2D setting as that in the last example is applied. The far-field predictor, $u_0$, is given in Appendix~\ref{sec:sub:apd:disloc}. Figure~\ref{fig:geom_crack} illustrates the corresponding QM/MM decomposition and the core geometry. Again, we choose a sufficiently large computational domain and MM region, $R_{\Omega}=100r^{\rm Si}_0$ and $\RMM=90r^{\rm Si}_0$, where $r^{\rm Si}_0$ is the lattice constant of Si (diamond).

\begin{figure}[!htb]
    \centering
    \includegraphics[height=5cm]{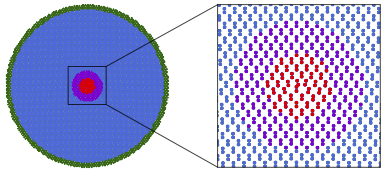}
    \caption{Illustration of (110)[100] edge dislocation in Si.}
    \label{fig:geom_edge}
\end{figure}

The construction of the ACE potential used for QM/MM models for edge dislocation in Si is the same as that for edge dislocation in W, given in Section \ref{sec:sub:params_study}. The only difference is that the data in training set is evaluated by NRL-TB model instead of EAM potential. The {\it a priori} error estimate of corresponding QM/MM models for edge dislocation reads~\cite[Theorem 3.4]{2021-qmmm3}
$$\|\bar{u} - \uH\|_{\UsH} \lesssim N_{\rm QM}^{-1.0}.$$ 

\begin{figure}[!htb]
    \centering
    \includegraphics[height=6cm]{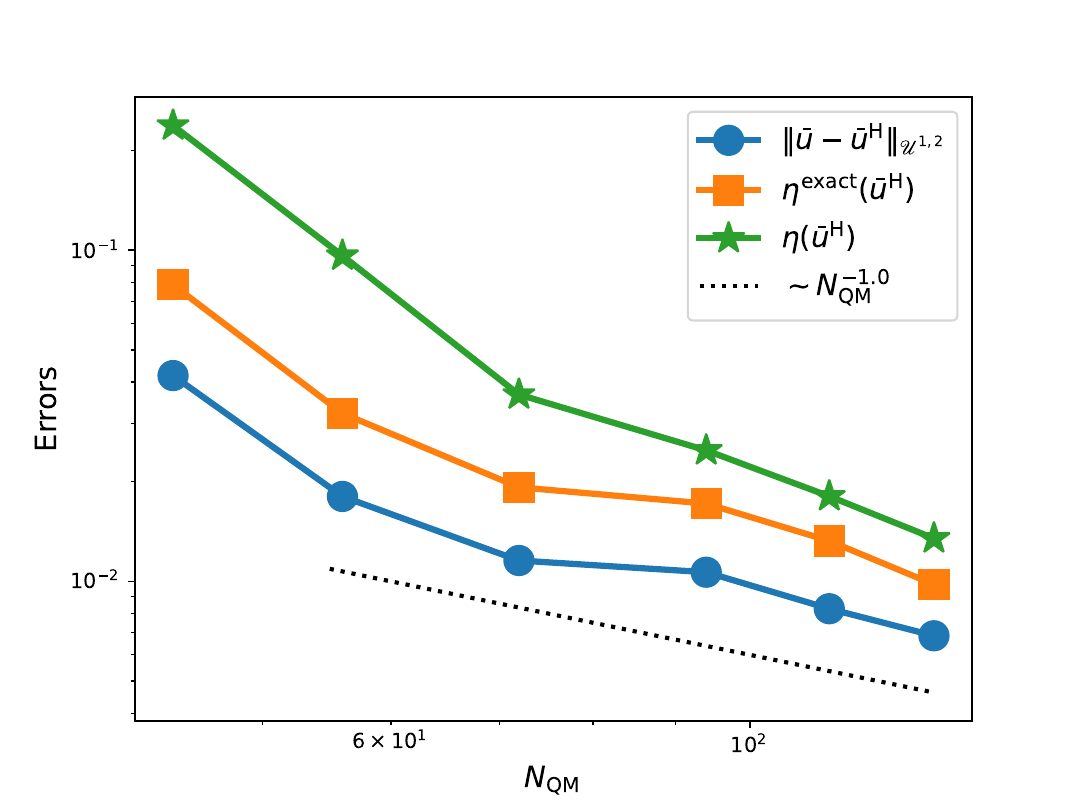}
    \caption{NRL-TB Si model: Convergence of the adaptive Algorithm \ref{alg:main} for (110)[100] edge dislocation.}
    \label{fig:conv_disloc_Si}
\end{figure}

Figure \ref{fig:conv_disloc_Si} plots the convergences of the approximation error $\|\bar{u}-\uH\|_{\UsH}$, and the error estimators $\eta(\uH)$, $\eta^{\rm exact}(\uH)$ against $N_{\rm QM}$ during the adaptive computations. Similar to the Figure \ref{fig:conv_crack} shown in the last section, we observe that two error estimators follow the trend of the approximation error fairly closely, which verifies again that $\eta(\uH)$ and $\eta^{\rm exact}(\uH)$ can provide efficient and reliable estimators for NRL-TB Si is the reference model. Moreover, the difference between $\eta(\uH)$ and $\eta^{\rm exact}(\uH)$ is still marginal (with a prefactor 1.51) especially when $N_{\rm QM}$ is large, which demonstrates the accuracy of the {\it practical} error estimator. 

\begin{figure}[!htb]
    \centering
    \includegraphics[height=6cm]{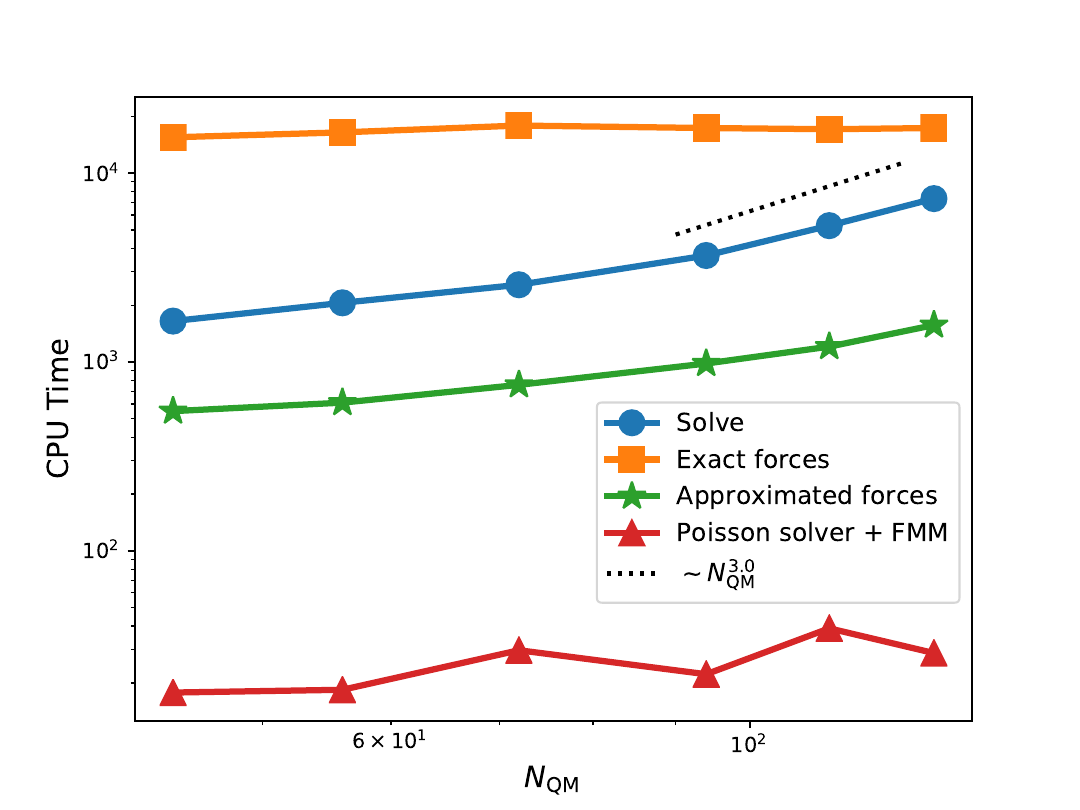}
    \caption{NRL-TB Si model: CPU time (s) of different components (steps) in Algorithm \ref{alg:main} for (110)[100] edge dislocation.}
    \label{fig:time_disloc_Si}
\end{figure}

Figure~\ref{fig:time_disloc_Si} shows the CPU time (in seconds) of running different components (steps) in Algorithm~\ref{alg:main}. As an actual QM (NRL-TB) model is used as the reference model, the cost of computing the exact residual forces is extremely expensive, which requires to solve an eigenvalue problem on the whole computational domain $\Omega$, even exceeds that of solving QM/MM solution $\uH$. The scaling of solving (blue line) is nearly cubic asymptotically, confirming that the computational cost of solving for $\uH$ is about $O(N^3_{\rm QM})$ as the cost to solve the QM (NRL-TB) model scales cubically. We also note that the costs of solving~\eqref{eq:stress_force_trun_T} for the error estimator and solving~\eqref{eq:eikonal} for the first arriving time are both negligible compare to others. More importantly, the evaluation time of computing the approximated forces $\widetilde{\F}(\uH)$ defined by~\eqref{eq:ACE2F} is significantly reduced compared with that of evaluating the exact residual forces $\F(\uH)$ while the accuracy can still be retained (cf. Figure~\ref{fig:conv_disloc_Si}). This observation verifies the accuracy and efficiency of our main adaptive algorithm (Algorithm~\ref{alg:main}). 

\begin{figure}[!htb]
    \centering
    \includegraphics[height=3.75cm]{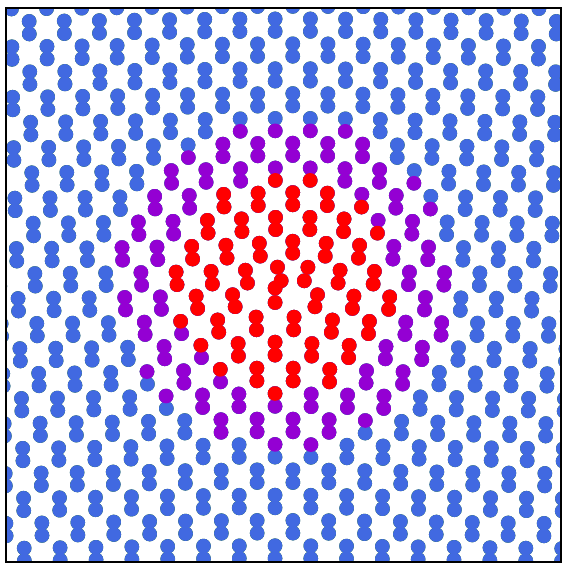}
    \includegraphics[height=3.75cm]{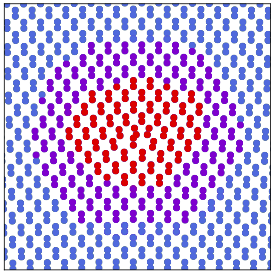}
    \includegraphics[height=3.75cm]{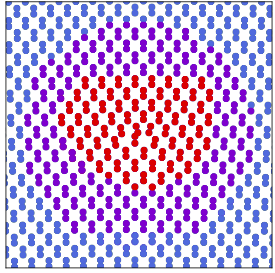}
    \includegraphics[height=3.75cm]{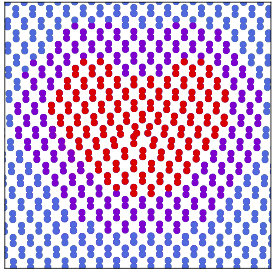}
    \caption{Evolution of QM and MM partitions in the adaptive procedure for (110)[100] edge dislocation in Si. $N_{\rm QM}$ are 76, 94, 110, 136 from left to right.}
    \label{fig:evolution_disloc_Si}
\end{figure}

The evolution of the QM/MM partitions during the adaptation process is shown in Figure~\ref{fig:evolution_disloc_Si}. We observe that the adaptive algorithm adjusts the QM region anisotropically during the adaptive computations, which demonstrates the robustness of our main algorithm (Algorithm~\ref{alg:main}) when an actual QM model is considered.

\subsection{Di-interstitial in Si}
\label{sec:sub:numerics_diinterstitial}

\begin{figure}[!htb]
    \centering
    \includegraphics[height=6cm]{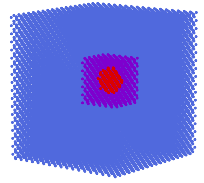}
    \caption{Domain decomposition for di-interstitial in Si. No far-field (FF) region due to the periodic boundary conditions are used in all directions.}
    \label{fig:geom_diint}
\end{figure}

The last type of defect we consider is the di-interstitial in Si in three dimensions, where periodic boundary conditions are used in all directions. The construction of a di-interstitial in Si follows from~\cite{bartok2018machine}. The illustration of its QM/MM decomposition is given in Figure~\ref{fig:geom_diint}. The computational domain includes $10^3$ supercells, containing 8002 Si atoms. We first choose NRL-TB as the reference model, and then perform our main adaptive algorithm for a more realistic QM model, where the plane-wave DFT~\cite{lin2019mathematical, martin2020electronic} is applied. 

To construct the ACE potential for the QM/MM model for the di-interstitial the training set contains only the QM force constant ($\delta f({\bf 0})$) at the homogeneous lattice $\Lhom$. The ACE potential is constructed by directly minimizing~\eqref{cost:forcemix}. The corresponding QM/MM model has the following {\it a priori} error estimate~\cite[Theorem 3.3]{2021-qmmm3}  
$$\|\bar{u} - \uH\|_{\UsH} \lesssim N_{\rm QM}^{-1.5}.$$

\subsubsection{NRLTB model}

\begin{figure}[!htb]
    \centering
    \includegraphics[height=6cm]{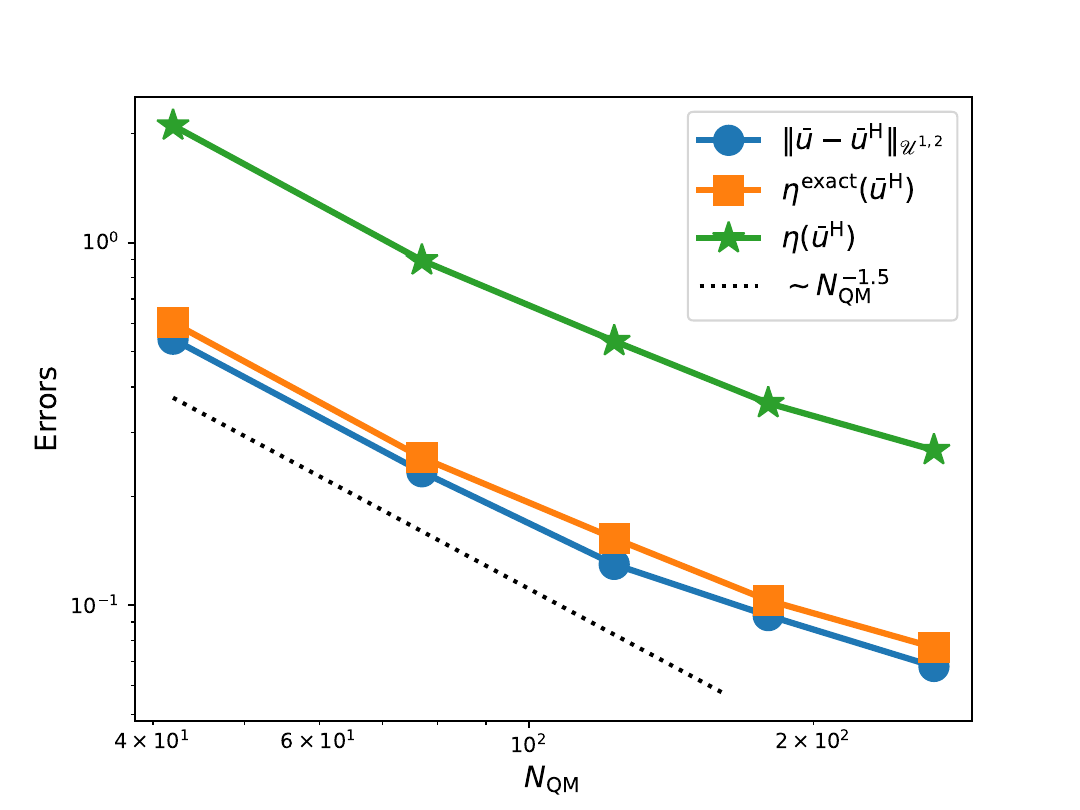}
    \caption{NRL-TB Si model: Convergence of the adaptive Algorithm \ref{alg:main} for di-interstitial.}
    \label{fig:conv_diint_Si}
\end{figure}

We first choose NRLTB model as the reference QM model to study the convergence of our main adaptive algorithm. We plot the approximation error $\|\bar{u}-\uH\|_{\UsH}$ and the error estimators $\eta(\uH)$, $\eta^{\rm exact}(\uH)$ against $N_{\rm QM}$ in Figure~\ref{fig:conv_diint_Si}. The result again shows that both two error estimators are efficient and reliable and $\eta(\uH)$ provides an accurate approximation (with a prefactor 1.87). Note that the final data point in Figure~\ref{fig:conv_diint_Si} exhibits sub-optimal behavior since the relationship between $N_{\rm MM}$ and $N_{\rm QM}$ for balancing the approximation error ($N_{\rm MM}\approx cN_{\rm QM}^{3.0}$ with a constant $c$), is no longer satisfied~\cite[Theorem 3.3]{2021-qmmm3}. We anticipate that this issue can be addressed by selecting a larger MM region instead.

\begin{figure}[!htb]
    \centering
    \includegraphics[height=6cm]{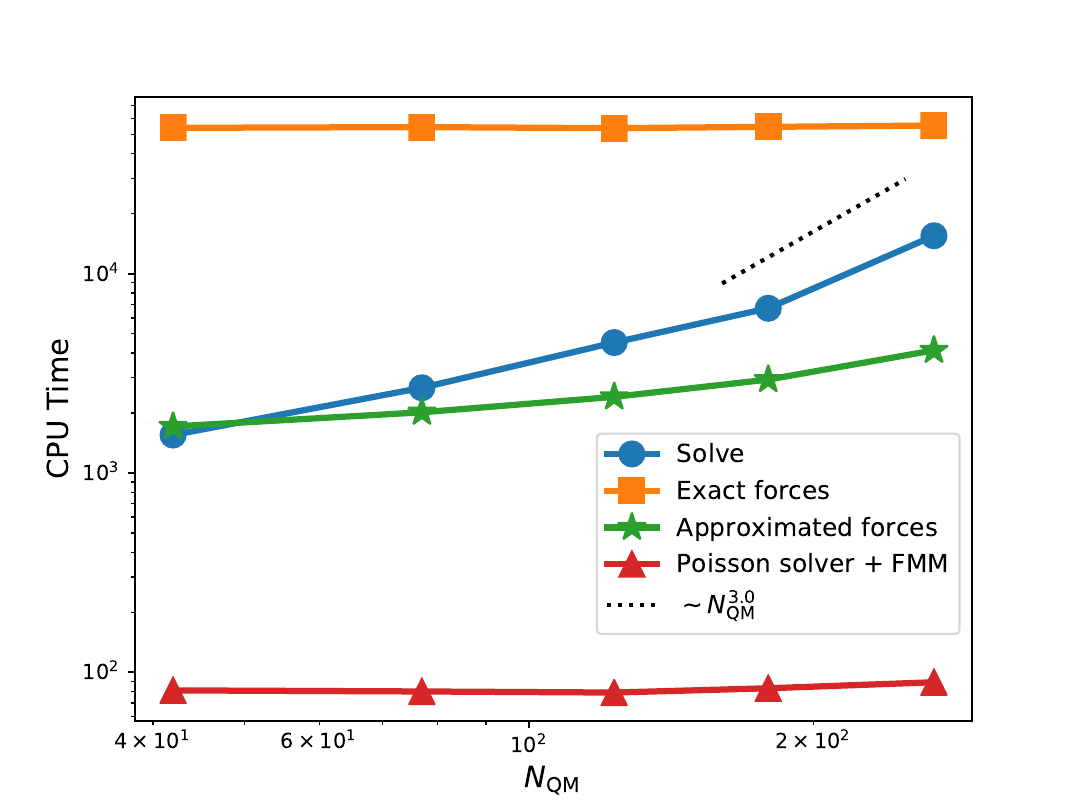}
    \caption{NRL-TB Si model: CPU time (s) of different components (steps) in Algorithm \ref{alg:main} for di-interstitial.}
    \label{fig:time_diint_Si}
\end{figure}

We also show the corresponding CPU time (in seconds) of running different components in Algorithm~\ref{alg:main} in Figure~\ref{fig:time_diint_Si}. Again it is clear to see the evaluation time of computing the approximated forces $\widetilde{\F}(\uH)$ defined by~\eqref{eq:ACE2F} is significantly reduced compared with that of evaluating the exact residual forces $\F(\uH)$, which demonstrates the efficiency of the {\it practical} error estimator. 

\begin{figure}[!htb]
    \centering
    \includegraphics[height=3.8cm]{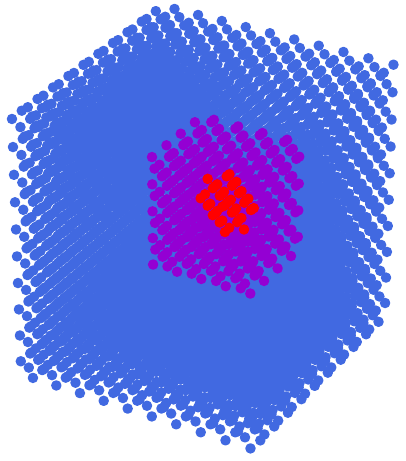}
    \includegraphics[height=3.8cm]{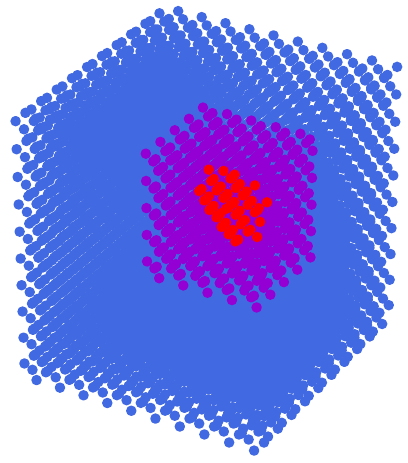}
    \includegraphics[height=3.8cm]{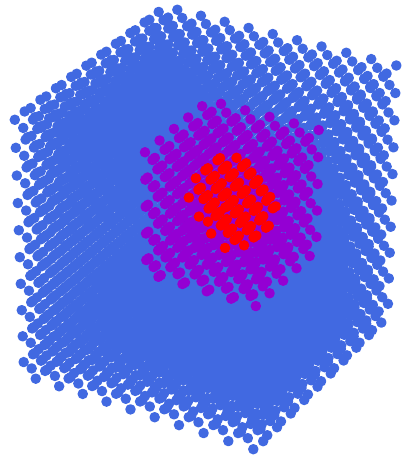}
    \includegraphics[height=3.8cm]{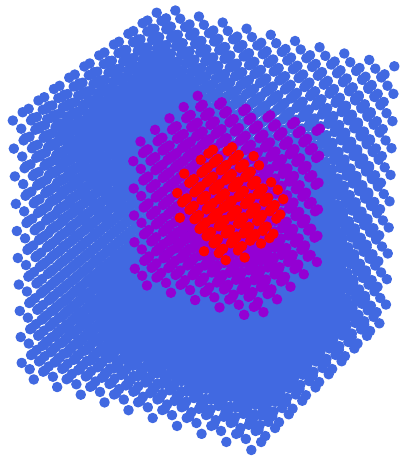}
    \caption{Evolution of QM and MM partitions in the adaptive procedure for di-interstitial in Si. $N_{\rm QM}$, the numbers of atomic sites in the QM region are 42, 77, 123, 179 from left to right. Note that the defect core is not placed at the origin.}
    \label{fig:evolution_diint_Si}
\end{figure}

The evolution of the QM/MM partitions during the adaptation process is presented in Figure~\ref{fig:evolution_diint_Si}.
In this scenario, the buffer region is extended cubically, thereby maintaining isotropic extension, to ensure compatibility with the plane-wave DFT calculation discussed in the subsequent section. 

\subsubsection{DFT model}
\label{sec:sub:sub:dft}

We still consider the di-interstitial case but now choosing plane-wave DFT as the reference model. In this setting we do not have rigorous locality results~\assERL~as for the NRLTB model, but we still expect that some (unknown) variation of those results remains true. Our interest is therefore to explore 
whether the main adaptive algorithm (Algorithm~\ref{alg:main}) can be applied in this setting as well. We utilize an open-source {\tt Julia} package {\tt DFTK.jl}~\cite{herbst2021dftk}. The main parameters chosen are as follows: plane-wave cutoff 200eV, $\Gamma$-centered $k$ meshes with $4\times 4 \times 4$ $k$-points, 0.1eV smearing with Fermi-Dirac smearing method.

The ACE potential for DFT is constructed in a similar way as that for NRL-TB, the only difference is that we consider more random configurations in the training set in order to obtain an accurate ACE potential from DFT calculations.

\begin{figure}[!htb]
    \centering
    \includegraphics[height=6cm]{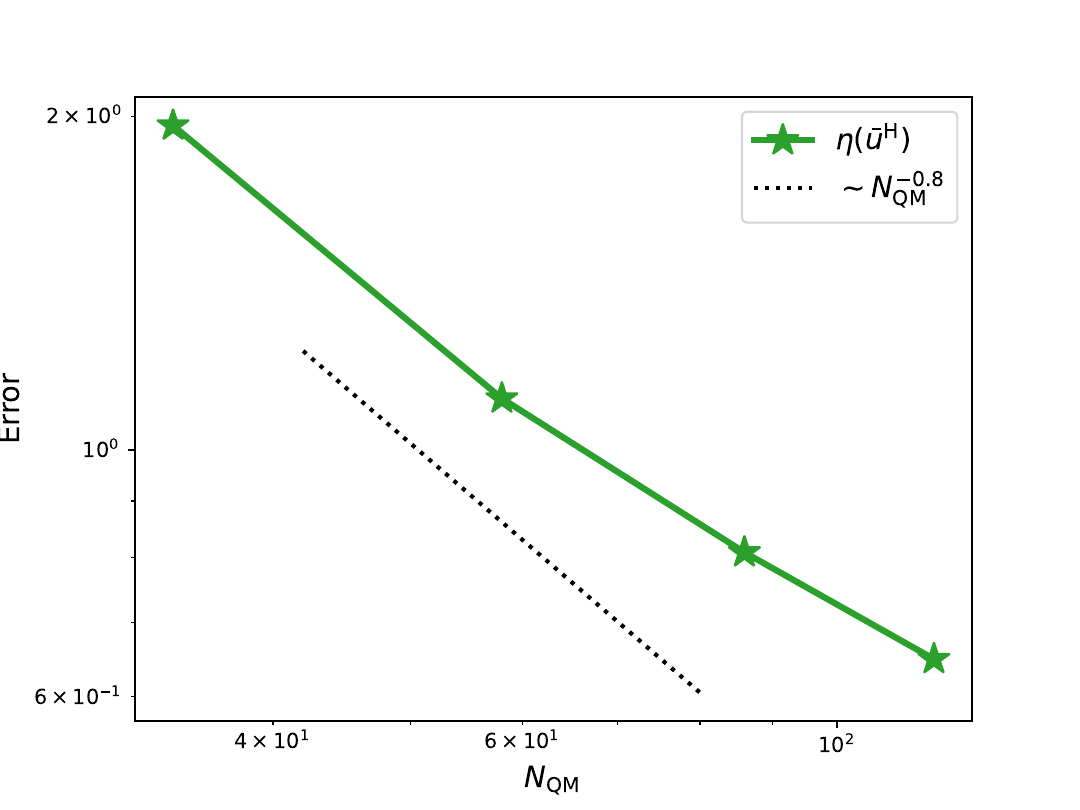}
    \caption{DFT Si model: Convergence of of the adaptive Algorithm \ref{alg:main} for di-interstitial.}
    \label{fig:conv_diint_Si_DFT}
\end{figure}

Since the reference solution $\bar{u}$ and the exact residual forces can not be computed in practice due to highly expensive computational cost, we only plot the {\it approximated} error estimator $\eta(\uH)$ against $N_{\rm QM}$ in Figure~\ref{fig:conv_diint_Si_DFT}. A sup-optimal convergence rate ($N^{-0.8}_{\rm QM}$) is observed, and we explain that it probably comes from the error in the buffer region due to insufficient buffer region size. Although the optimal convergence can not be achieved temporarily, our main adaptive algorithm can still work for DFT simulation, which is already a significant improvement in the field of adaptive QM/MM methods. A detailed study for practical DFT/MM models, including how to apply extrapolation technique to obtain the reference solutions for DFT, will be investigated in our future work.

\section{Conclusion}
\label{sec:conclusion}

We proposed a novel adaptive QM/MM method for practical material defect simulations. To ensure  {\it consistency} of the QM/MM method with the reference QM model, we employ ``machine-learned interatomic potentials (MLIPs)" as the MM models~\cite{2021-qmmm3}. Our adaptive QM/MM method utilizes a residual-based error estimator that provides both upper and lower bounds for the approximation error, thus indicating its reliability and efficiency. Furthermore, we introduce a novel adaptive algorithm capable of anisotropically updating the QM/MM partitions. This update is based on the proposed residual-based error estimator and involves solving a free interface motion problem, which is efficiently addressed using the fast marching method. To demonstrate the robustness of our approach, we performed numerical simulations involving a range of crystalline defects (point defects, dislocations, cracks). 

Our results suggest that the proposed adaptive algorithm is generally applicable for other common multiscale coupling schemes and more complex crystalline defects, some open problems remain that deserve further mathematical and empirical analysis as well as further algorithmic developments, for example:

{\it More complex crystalline defects:} More complex defect structures such as two partial dislocations connected by a stacking fault, or dislocation nucleation, are mcuh more difficult to include in a rigorous mathematical analysis, but we see no reason why our methods are not applicable in principle. Indeed, such scenarios are exactly where an adaptive QM/MM method can demonstrate the maximum gain.

{\it Quasi-static and dynamical problems:} We plan to explore the generalization of this work to adaptive error control for quasi-static and dynamical problems (with moving defects). Our method is potentially more efficient and important for these problems, where both model refinement and coarsening must be carefully considered. This would require some adaptations to our methodology.  


\appendix

\section{Proofs}
\label{sec:proofs}

In this section, we give the rigorous proofs of the main results in Section~\ref{sec:errest} (cf. Lemma~\ref{lemma:res-F} and Theorem~\ref{th:ctsphi}) and present the sketch of proof for the {\it a priori} error estimate of QM/MM coupling for crack (cf. \eqref{eq:apriori_crack}). 

\subsection*{Proof of Lemma~\ref{lemma:res-F}}
Following the analysis in \cite[Lemma 3.1]{CMAME} and extending it to the force-mixing scheme by applying the techniques in \cite[Appendix C]{chen17}, we can prove Lemma~\ref{lemma:res-F}.
\begin{proof}
Suppose $\bar{u}$ is a strongly stable solution of \eqref{eq:problem-force}, applying the fact that $\<\mathcal{F}(\bar{u}), v\>=0$ for any $v\in \Us^{1,2}$ and the Lipschitz continuity of $\F$ \cite{chen19, chen16}, we have
\[
\< \F(\uH) - \F(\bar{u}), v\> \leq C \|\bar{u} - \uH\|_{\UsH} \|v\|_{\UsH}, \quad \forall v\in \Us^{1,2},
\]
which leads to the lower bound of the true approximation error
\begin{eqnarray}\label{eq:lo}
\| \F(\uH) \|_{(\UsH)^*} \leq C\|\bar{u}-\uH\|_{\UsH}.
\end{eqnarray}
Combining the analogous techniques in the proof of \cite[Lemma 3.1]{CMAME} with the estimate in \cite[Appendix C]{chen17}, we can obtain the upper bound
\begin{eqnarray}\label{eq:up}
\| \F(\uH) \|_{(\UsH)^*} \geq c\|\bar{u}-\uH\|_{\UsH}.
\end{eqnarray}
Taking into account the results \eqref{eq:lo} and \eqref{eq:up}, we can yield the stated results.
\end{proof}

\subsection*{Proof of Theorem~\ref{th:ctsphi}}

Theorem~\ref{th:ctsphi} demonstrates that $\phi_{\a}$ serves as a Riesz representation for the residual $\F(\uH)$. This approach, in contrast to our previous method~\cite{wang2020posteriori}, proves to be both practical and versatile by substituting the PDE operator with a generalized form of the Laplacian matrix.

\begin{proof}
To simplify notation, we use $\lesssim$ to indicate $\leq C$, where $C$ is a constant independent of model parameters. Additionally, $\eqsim$ denotes both $\lesssim$ and $\gtrsim$. Given the definition of a generalization of the Laplacian matrix used to represent undirected graphs by \eqref{LMatrix}, it is straightforward to see that for any $u\in \UsH$,
\[
\langle L^{\L} u, u \rangle \eqsim \| D u \|^2_{\ell^2_{\mathcal{N}}(\L)}.
\]
Given the residual force $\F(\uH)$, by duality, one can derive
\[
\big\langle (L^{\L})^{-1} \F(\uH), \F(\uH) \big\rangle \eqsim \| \F(\uH) \|_{(\UsH)^*}.
\]
Since $\phi_{\a}(\uH)$ solves the equation \eqref{eq:Lphi}, i.e., $L^{\L} \cdot \phi_{\a}(\uH) = \F(\uH)$, we have 
\[
\| D \phi_{\a}(\uH) \|^2_{\ell^2_{\mathcal{N}}(\L)} = \| D (L^{\L})^{-1} \F(\uH) \|^2_{\ell^2_{\mathcal{N}}(\L)} \eqsim \big\langle L^{\L} (L^{\L})^{-1} \F(\uH), (L^{\L})^{-1} \F(\uH) \big\rangle \eqsim \| \F(\uH) \|_{(\UsH)^*}.
\]
This establishes the estimate \eqref{eq:equiv-1}. Combining this error estimate with \eqref{res-bound}, we can obtain \eqref{eq:equiv-2}, which completes the proof.
\end{proof}

\def\uh{u_0}

\subsection*{Proof of the estimate \eqref{eq:apriori_crack}}
Here we provide a sketch of the proof of the estimate \eqref{eq:apriori_crack}. We admit that the details including the analysis of the elastic field induced by crack geometries as well as the stability analysis, appear to be considerably challenging, which will be investigated rigorously in our future work. See \cite{2019-crackbif} for some recent advances in this direction.
\begin{proof}
The main idea is to adapt the proof of \cite[Theorem 3.4]{2021-qmmm3} to crack by employing the fact that the far-field predictor for crack decays as $|D u_0(\ell)|\lesssim |\ell|^{-0.5}$ \cite{2019-crackbif}.
According to the proof shown in \cite[Appendix A]{2021-qmmm3}, for any $v \in \Admu^{\rm H}$, we have the following consistency estimate 
	\begin{align}\label{e-mix-consistency-v}
		\big\< \mathcal{F}^{\rm H}(T^{\rm H}\bar{u}), v \big\> \lesssim & 
		\Big( \sum^{K}_{j=1} \ffit_{j} \|D\uh\|^{j}_{\ell^{2j}_{\wf}(\LMM\cup\LFF)} + \vfit_{K+1} \|\nabla \uh\|^{K+1}_{L^{2K+2}(\LMM \cup \LFF)}
		\nonumber \\
		&+ \|\nabla \uh\|^{K}_{L^{2K}(\LMM \cup \LFF)}\|\nabla^2 \uh\|_{L^2(\LMM \cup \LFF)}
		\Big) \cdot \|Dv\|_{\ell^2_{\mathcal{N}}},
	\end{align}
where $T^{\rm H}$ is the mapping from $\Admu$ to $\Admu^{\rm H}$. In our setting, we choose $K=1$. Hence, we can roughly obtain the estimate \eqref{eq:apriori_crack} by combining the decay estimate for crack with a suitable stability analysis. 
\end{proof}

\section{Far-field predictors}
\label{sec:appendixU0}

\subsection{Dislocation}
\label{sec:sub:apd:disloc}

\newcommand{\ulin}{u^{\rm lin}}
\newcommand{\burg}{{\sf b}}

A model for straight dislocations is considered in this work, following the setting proposed in \cite{Ehrlacher16}. Specifically, the model is constructed by projecting a three-dimensional crystal onto a two-dimensional plane. Let $B \in \R^{3\times 3}$ be a nonsingular matrix. Given a Bravais lattice $B\Z^3$ with dislocation direction parallel to $e_3$ and Burgers vector $\burg=(\burg_1,0,\burg_3)$, we consider
displacements $W: B\Z^3 \rightarrow \R^3$ that are periodic in the direction of the dislocation direction of $e_3$. Thus, we choose a projected reference lattice $\L := A\Z^2 := \{(\ell_1, \ell_2) ~|~ \ell=(\ell_1, \ell_2, \ell_3) \in B\Z^3\}$. We also introduce the projection operator 
\begin{equation}\label{eq:P}
    P(\ell_1, \ell_2) = (\ell_1, \ell_2, \ell_3) \quad \text{for}~\ell \in B\Z^3.
\end{equation}
It can be readily checked that this projection is again a Bravais lattice.

We prescribe the far-field predictor $u_0$ as follows according to \cite{chen19, Ehrlacher16}. Let $\L\subset\R^2$, $\hat{x}\in\R^2$ be the position of the dislocation core and $\Gamma := \{x \in \R^2~|~x_2=\hat{x}_2,~x_1\geq\hat{x}_1\}$ be the ``branch cut'', with $\hat{x}$ chosen such that $\Gamma\cap\Lambda=\emptyset$.
We define the far-field predictor $u_0$ by
\begin{eqnarray}\label{predictor-u_0-dislocation}
u_0(x):=\ulin(\xi^{-1}(x)),
\end{eqnarray}
where $\ulin \in C^\infty(\R^2 \setminus \Gamma; \R^d)$ is the solution of continuum linear elasticity (CLE)
\begin{align}\label{CLE}
\nonumber
\mathbb{C}^{j\beta}_{i\alpha}\frac{\partial^2 u^{\rm lin}_i}{\partial x_{\alpha}\partial x_{\beta}} = 0 \qquad &\text{in} ~~ \R^2\setminus \Gamma,
\\
u^{\rm lin}(x+) - u^{\rm lin}(x-) = -\burg \qquad &\text{for} ~~  x\in \Gamma \setminus \{\hat{x}\},
\\
\nonumber
\nabla_{e_2}u^{\rm lin}(x+) - \nabla_{e_2}u^{\rm lin}(x-) = 0 \qquad &\text{for} ~~  x\in \Gamma \setminus \{\hat{x}\},
\end{align}
where the forth-order tensor $\mathbb{C}$ is the linearised Cauchy-Born tensor (derived from the potential $\Vhom$, see \cite[\S~7]{Ehrlacher16} for more detail),
\begin{eqnarray}
\xi(x)=x-\burg_{12}\frac{1}{2\pi}
\eta\left(\frac{|x-\hat{x}|}{\hat{r}}\right)
\arg(x-\hat{x}),
\end{eqnarray}
with $\arg(x)$ denoting the angle in $(0,2\pi)$ between $x$ and
$\burg_{12} = (\burg_1, \burg_2) = (\burg_1, 0)$, and
$\eta\in C^{\infty}(\R)$ with $\eta=0$ in $(-\infty,0]$ and $\eta=1$ in
$[1,\infty)$ which removes the singularity. It is widely recognized that the gradient of the displacement field $u_0$ follows $r^{-1}$ with respect to the distance from $\hat{x}$.


\subsection{Crack}
\label{sec:sub:apd:crack}

We present the setting of crack by following \cite{2019-crackbif}, which stems from the limitation of the continuum elasticity approaches to static crack problems. Similar with the discussions of dislocations, we introduce the following CLE
\begin{align}\label{CLE_crack}
\nonumber
- {\rm div}~(\mathbb{C}:\nabla u) = 0 \qquad &\text{in} ~~ \R^2\setminus \Gamma,
\\
(\mathbb{C} : \nabla u)\nu = 0 \qquad &\text{on} ~~ \Gamma,
\end{align}
supplied with a suitable boundary condition coupling to the bulk \cite{freund1998dynamic}. It is well-known that in the vicinity of the crack tip, the gradients of solutions to \eqref{CLE_crack} exhibit a persistent $1/\sqrt{r}$ behaviour, where $r$
is the distance from the crack tip (cf. \cite{rice1968mathematical}). 


\section{The Atomic Cluster Expansion}
\label{sec:ACE}


Following \cite{bachmayr19}, we briefly introduce the construction of the ACE potential. 
Given $\mathcal{N} \in \N$, we first write the ACE site potential in the form of an {\it atomic body-order expansion}, $\displaystyle V^{\rm ACE} \big(\{\pmb{g_j}\}\big) = \sum_{N=0}^{\mathcal{N}} \frac{1}{N!}\sum_{j_1 \neq \cdots \neq j_N} V_N(\pmb{g}_{j_1}, \cdots, \pmb{g}_{j_N})$, where the $N$-body potential $V_N : \R^{dN} \rightarrow \R$ can be approximated by using a tensor product basis \cite[Proposition 1]{bachmayr19},
\begin{align*}
\phi_{\pmb{k\ell m}}\big(\{\pmb{g}_j\}_{j=1}^N\big) :=\prod_{j=1}^N\phi_{k_j\ell_j m_j}(\pmb{g}_j) 
\quad & {\rm with}\quad 
\phi_{k\ell m}(\pmb{r}):=P_k(r)Y^{m}_{\ell}(\hat{r}) , ~~\pmb{r}\in\R^d,~r=|\pmb{r}|,~\hat{r}=\pmb{r}/r ,
\end{align*}
where $P_k,~k=0,1,2,\cdots$ are radial basis functions (for example, Jacobic polynomials), and $Y_{\ell}^m,~\ell = 0,1,2,\cdots,~m=-\ell,\cdots,\ell$ are the complex spherical harmonics.
The basis functions are further symmetrised to a permutation invariant form,
\begin{eqnarray*}
\tilde{\phi}_N = \sum_{(\pmb{k,\ell,m})~{\rm ordered}} \sum_{\sigma\in S_N} \phi_{\pmb{k\ell m}} \circ \sigma ,
\end{eqnarray*}  
where $S_N$ is the collection of all permutations, and by $\sum_{(\pmb{k,\ell,m})~{\rm ordered}}$ we mean that the sum is over all lexicographically ordered tuples $\big( (k_j, \ell_j, m_j) \big)_{j=1}^N$.
The next step is to incorporate the invariance under point reflections and rotations
\begin{eqnarray*}
\mathcal{B}_{\pmb{k\ell} i} = \sum_{\pmb{m}\in\mathcal{M}_{\pmb{\ell}}}\mathcal{U}_{\pmb{m}i}^{\pmb{k\ell}} \sum_{\sigma\in S_N} \phi_{\pmb{k\ell m}} \circ \sigma 
\quad{\rm with}\quad \mathcal{M}_{\pmb{\ell}} = \big\{\pmb{\mu}\in\Z^N ~|~ -\ell_{\alpha}\leq\mu_{\alpha}\leq\ell_{\alpha} \big\} ,
\end{eqnarray*}
where the coefficients $\mathcal{U}_{mi}^{k\ell}$ are given in \cite[Lemma 2 and Eq. (3.12)]{bachmayr19}.
It was shown in \cite{bachmayr19} that the basis defined above is explicit but computational inefficient. The so-called ``density trick" technique used in \cite{Bart10, Drautz19, Shapeev16} can transform this basis into one that is computational efficient. The alternative basis is 
\begin{eqnarray*}
B_{\pmb{k\ell} i} = \sum_{\pmb{m}\in\mathcal{M}_{\pmb{\ell}}}\mathcal{U}_{\pmb{m}i}^{\pmb{k\ell}} A_{\pmb{nlm}}
\quad{\rm with~the~correlations}\quad A_{\pmb{nlm}}:= \prod_{\alpha=1}^{N} \sum_{j=1}^{J} \phi_{n_\alpha l_\alpha m_{\alpha}}(\pmb{g}_j),
\end{eqnarray*}
which avoids both the $N!$ cost for symmetrising the basis as well as the $C_J^N$ cost of summation over all order $N$ clusters within an atomic neighbourhood.
The resulting basis set is then defined by
\begin{eqnarray}
\pmb{B}_N := \big\{ B_{\pmb{k\ell} i}  ~|~ (\pmb{k},\pmb{\ell})\in\N^{2N}~{\rm ordered}, ~\sum_{\alpha} \ell_{\alpha}~{\rm even},~ i=1,\cdots,\pmb{n}_{\pmb{k\ell}}\big\},
\end{eqnarray}
where $\pmb{n}_{\pmb{k\ell}}$ is the rank of body-orders (see \cite[Proposition 7 and Eq. (3.12)]{bachmayr19}).

Once the finite symmetric polynomial basis set $\pmb{B} \subset \bigcup^{\mathcal{N}}_{N=1} \pmb{B}_N$ is constructed, the ACE site potential can be expressed as
\begin{align}\label{ships:energy}
V^{\rm ACE}(\pmb{g}; \{c_B\}_{B\in\pmb{B}}) = \sum_{B\in\pmb{B}} c_B B(\pmb{g}) 
\end{align}
with the coefficients $c_{B}$. The corresponding force of this potential is denoted by $\F^{\rm ACE}$.

The family of potentials are systematically improvable (see \cite[Section 6.2]{bachmayr19}): by increasing the body-order, cutoff radius and polynomial degree they are in principle capable of representing an arbitrary many-body potential energy surface to within arbitrary accuracy~\cite{witt2023acepotentials}.

\section{A semi-empirical QM model: The NRL tight binding}
\label{sec:NRL}

In this paper, we use the tight binding model as the reference quantum mechanical model for simplicity of presentation. We note that our numerical scheme is in principle also suitable for general quantum mechanical models. 

The NRL tight binding model is developed by Cohen, Mehl, and Papaconstantopoulos \cite{cohen94}. The energy  levels are determined by the generalised eigenvalue problem
\begin{eqnarray}\label{NRL:diag_Ham}
\mathcal{H}(y)\psi_s = \lambda_s\mathcal{M}(y)\psi_s 
\qquad\text{with}\quad \psi_s^{\rm T}\mathcal{M}(y)\psi_s = 1,
\end{eqnarray}
where $\mathcal{H}$ is the hamiltonian matrix and $\mathcal{M}(y)$ is the overlap matrix. The NRL hamiltonian and overlap matrices are construct both from hopping elements as well as on-site matrix elements as a function of the local environment. For carbon and silicon they are parameterised as follows (for other elements the parameterisation is similar): 

To define the on-site terms, each atom $\ell$ is assigned a pseudo-atomic density 
\begin{eqnarray*}
	\rho_{\ell} := \sum_{k}e^{-\lambda^2 r_{\ell k}}  \fc(r_{\ell k}),
\end{eqnarray*}
where the sum is over all of the atoms $k$ within the cutoff $\Rc$ of atom $\ell$, $\lambda$ is a fitting parameter, $\fc$ is a cutoff function
\begin{eqnarray*}
	\fc(r) = \frac{\theta(\Rc-r)}{1+\exp\big((r-\Rc)/l_c + L_c\big)} ,
\end{eqnarray*}
with $\theta$ the step function, and the parameters $l_c=0.5$, $L_c=5.0$ for most elements.
Although, in principle, the on-site terms should have off-diagonal elements, but this would lead to additional computational challenges that we wished to avoid. The NRL model follows traditional practice and only include the diagonal terms.
Then, the on-site terms for each atomic site $\ell$ are given by
\begin{eqnarray}\label{NRLonsite}
\mathcal{H}(y)_{\ell\ell}^{\upsilon\upsilon}
:= a_{\upsilon} + b_{\upsilon}\rho_{\ell}^{2/3} + c_{\upsilon}\rho_{\ell}^{4/3} + d_{\upsilon}\rho_{\ell}^{2},
\end{eqnarray}
where $\upsilon=s,p$, or $d$ is the index for angular-momentum-dependent atomic orbitals and $(a_\upsilon)$, $(b_\upsilon)$, $(c_\upsilon)$, $(d_\upsilon)$ are fitting parameters.  The on-site elements for the overlap matrix are simply taken to be the identity matrix.

The off-diagonal NRL Hamiltonian entries follow the formalism of  Slater and Koster who showed in \cite{Slater54} that all two-centre (spd) hopping integrals can be constructed from ten independent ``bond integral'' parameters $h_{\upsilon\upsilon'\mu}$, where
\begin{eqnarray*}
	(\upsilon\upsilon'\mu) = ss\sigma,~sp\sigma,~pp\sigma,~pp\pi,
	~sd\sigma,~pd\sigma,~pd\pi,~dd\sigma,~dd\pi,~{\rm and}~dd\delta.
\end{eqnarray*}
The NRL bond integrals are given by
\begin{eqnarray}\label{NRLhopping}
h_{\upsilon\upsilon'\mu}(r) 
:= \big(e_{\upsilon\upsilon'\mu} + f_{\upsilon\upsilon'\mu}r + g_{\upsilon\upsilon'\mu} r^2 \big) e^{-h_{\upsilon\upsilon'\mu}r} \fc(r)
\end{eqnarray}
with fitting parameters $e_{\upsilon\upsilon'\mu}, f_{\upsilon\upsilon'\mu},  g_{\upsilon\upsilon'\mu}, h_{\upsilon\upsilon'\mu}$. The matrix elements $\mathcal{H}(y)_{\ell k}^{\upsilon\upsilon'}$ are constructed from the $h_{\upsilon\upsilon'\mu}(r)$ by a standard procedure~\cite{Slater54}.

The analogous bond integral parameterisation of the overlap matrix 
is given by  
\begin{eqnarray}\label{NRLhopping-M}
m_{\upsilon\upsilon'\mu}(r) 
:= \big(\delta_{\upsilon\upsilon'} + p_{\upsilon\upsilon'\mu} r + q_{\upsilon\upsilon'\mu} r^2 + r_{\upsilon\upsilon'\mu} r^3 \big) e^{-s_{\upsilon\upsilon'\mu} r} \fc(r)
\end{eqnarray}
with the fitting parameters $(p_{\upsilon\upsilon'\mu}), (q_{\upsilon\upsilon'\mu}), (r_{\upsilon\upsilon'\mu}), (s_{\upsilon\upsilon'\mu})$ and $\delta_{\upsilon\upsilon'}$ the Kronecker delta function. 

The fitting parameters in the foregoing expressions are determined by fitting to some high-symmetry first-principle calculations: In the NRL method, a database of eigenvalues (band structures) and total energies were constructed for several crystal structures at  several volumes. Then the parameters are chosen such that the eigenvalues and energies in the database are reproduced.
For practical simulations, the parameters for different elements can be found in \cite{papaconstantopoulos15}.


\bibliographystyle{plain}
\bibliography{bib.bib}
\end{document}

%% file: notation.tex
\def\Z{\mathbb{Z}}
\def\a{\rm a}

\def\assERL{\textnormal{\bf (RL)}}

\def\Vhom{V^{\rm h}}

\def\E{\mathcal{E}}

\def\uh{u_{\rm h}}

\def\p0{\pmb{0}}

\def\Us{\mathscr{U}}
\def\uH{\bar{u}^{\rm H}}
\def\UsH{{\mathscr{U}}^{1,2}}

\def\ffit{\varepsilon^{\rm F}}
\def\vfit{\varepsilon^{\rm V}}
\def\L{\Lambda}
\def\R{\mathbb{R}}
\def\Rl{\mathcal{R}}
\def\Rg{\mathcal{R}^{*}}
\def\Adm{{\rm Adm}}
\def\<{\langle}
\renewcommand{\>}{\rangle}
\def\LQM{\Lambda^{\rm QM}}
\def\LMM{\Lambda^{\rm MM}}
\def\LFF{\Lambda^{\rm FF}}
\def\Lbuf{\Lambda^{\rm BUF}}
\def\Lhom{\L^{\rm h}}

\def\NQM{N_{\rm QM}}

\def\RMM{R_{\rm MM}}

\def\Nbuf{N_{\rm BUF}}
\def\Rbuf{R_{\rm BUF}}
\def\rcut{R_{\rm c}}
\def\Rcore{R_{\rm DEF}}

\def\F{\mathcal{F}}
\def\N{\mathbb{N}}

\def\Adm{{\rm Adm}}
\def\Admu{\mathscr{A}}
\def\wf{\mathfrak{w}}

\def\fc{f_{\rm c}}
\def\Rc{R_{\rm c}}

\def\Wcb{W_{\rm cb}}

\def\n{\mathfrak{n}}

%% file: GenAdapQMMM.bbl
\begin{thebibliography}{10}

\bibitem{bachmayr19}
M.~Bachmayr, G.~Csanyi, G.~Dusson, R.~Drautz, S.~Etter, C.~van~der Oord, and
  C.~Ortner.
\newblock Atomic cluster expansion: Completeness, efficiency and stability.
\newblock {\em J. Comp. Phys.}, 454:110946, 2022.

\bibitem{bartok2018machine}
A.~Bart{\'o}k, J.~Kermode, N.~Bernstein, and G.~Cs{\'a}nyi.
\newblock Machine learning a general-purpose interatomic potential for silicon.
\newblock {\em Phys. Rev. X}, 8(4):041048, 2018.

\bibitem{Bart10}
A.~Bart\'{o}k, M.~Payne, R.~Kondor, and G.~Cs\'{a}nyi.
\newblock Gaussian approximation potentials: {T}he accuracy of quantum
  mechanics, without the electrons.
\newblock {\em Phys. Rev. Let.}, 104:136403, 2010.

\bibitem{bernstein09}
N.~Bernstein, J.R. Kermode, and G.~Cs\'{a}nyi.
\newblock Hybrid atomistic simulation methods for materials systems.
\newblock {\em Rep. Prog. Phys.}, 72:26051 1--25, 2009.

\bibitem{boereboom2016}
J.~Boereboom, R.~Potestio, D.~Donadio, and R.~Bulo.
\newblock Toward hamiltonian adaptive qm/mm: accurate solvent structures using
  many-body potentials.
\newblock {\em J. Chem. Theory Comput.}, 12:3441--3448, 2016.

\bibitem{2019-crackbif}
M.~Buze, T.~Hudson, and C.~Ortner.
\newblock Analysis of cell size effects in atomistic crack propagation.
\newblock {\em ESAIM: Math. Model. Numer. Anal.}, 54:1821--1847, 2020.

\bibitem{CMAME}
H.~Chen, M.~Liao, H.~Wang, Y.~Wang, and L.~Zhang.
\newblock Adaptive {QM/MM} coupling for crystalline defects.
\newblock {\em Comput. Methods Appl. Mech. Engrg.}, 354:351--368, 2019.

\bibitem{chen18}
H.~Chen, J.~Lu, and C.~Ortner.
\newblock Thermodynamic limit of crystal defects with finite temperature tight
  binding.
\newblock {\em Arch. Ration. Mech. Anal.}, 230:701--733, 2018.

\bibitem{chen19}
H.~Chen, F.Q. Nazar, and C.~Ortner.
\newblock Geometry equilibration of crystalline defects in quantum and
  atomistic descriptions.
\newblock {\em Math. Models Methods Appl. Sci.}, 29:419--492, 2019.

\bibitem{chen16}
H.~Chen and C.~Ortner.
\newblock {QM/MM} methods for crystalline defects. {Part 1: L}ocality of the
  tight binding model.
\newblock {\em Multiscale Model. Simul.}, 14:232--264, 2016.

\bibitem{chen17}
H.~Chen and C.~Ortner.
\newblock {QM/MM} methods for crystalline defects. {Part 2: C}onsistent energy
  and force-mixing.
\newblock {\em Multiscale Model. Simul.}, 15:184--214, 2017.

\bibitem{chen19tb}
H.~Chen, C.~Ortner, and J.~Thomas.
\newblock Locality of interatomic forces in tight binding models for
  insulators.
\newblock {\em ESAIM: Math. Model. Numer. Anal.}, 54:2295--2318, 2020.

\bibitem{2021-qmmm3}
H.~Chen, C.~Ortner, and Y.~Wang.
\newblock Qm/mm methods for crystalline defects. part 3: Machine-learned
  interatomic potentials.
\newblock {\em ArXiv e-prints}, 2106.14559, 2021.

\bibitem{chopp2001some}
D.~Chopp.
\newblock Some improvements of the fast marching method.
\newblock {\em SIAM J. Sci. Comput.}, 23(1):230--244, 2001.

\bibitem{cohen94}
R.~Cohen, M.~Mehl, and D.~Papaconstantopoulos.
\newblock Tight-binding total-energy method for transition and noble metals.
\newblock {\em Phys. Rev. B}, 50:14694--14697, 1994.

\bibitem{csanyi04}
G.~Cs\'{a}nyi, T.~Albaret, M.~Payne, and A.~De Vita.
\newblock ``{L}earn on the fly": {A} hybrid classical and quantum-mechanical
  molecular dynamics simulation.
\newblock {\em Phys. Rev. Lett.}, 93:175503 1--4, 2004.

\bibitem{Daw1984a}
M.~Daw and M.~Baskes.
\newblock Embedded-atom method: {D}erivation and application to impurities,
  surfaces, and other defects in metals.
\newblock {\em Phys. Rev. B}, 29:6443--6453, 1984.

\bibitem{Dorfler:1996}
W.~D\"{o}rfler.
\newblock A convergent adaptive algorithm for poissons equation.
\newblock {\em SIAM J. Numer. Anal.}, 33:1106--1124, 1996.

\bibitem{Drautz19}
R.~Drautz.
\newblock Atomic cluster expansion for accurate and transferable interatomic
  potentials.
\newblock {\em Phys. Rev. B}, 99:014104, 2019.

\bibitem{duster17}
A.~Duster, C.~Wang, C.~Garza, D.~Miller, and H.~Lin.
\newblock Adaptive quantum/molecular mechanics: what have learned, where are
  we, and where do we go from here?
\newblock {\em WIREs Comput. Mol. Sci.}, 7:1--21, 2017.

\bibitem{Ehrlacher16}
V.~Ehrlacher, C.~Ortner, and A.~Shapeev.
\newblock Analysis of boundary conditions for crystal defect atomistic
  simulations.
\newblock {\em Arch. Ration. Mech. Anal.}, 222:1217--1268, 2016.

\bibitem{gitACEpotentials}
C.~Ortner {\it et al}.
\newblock {ACEpotentials.jl.git}.
\newblock https://github.com/ACEsuit/ACEpotentials.jl.

\bibitem{gitQMMM2}
C.~Ortner {\it et al}.
\newblock {QMMM2.jl.git}.
\newblock https://github.com:cortner/QMMM2.jl.git.

\bibitem{gitSKTB}
C.~Ortner {\it et al}.
\newblock {SKTB.jl.git}.
\newblock https://github.com/cortner/SKTB.jl.git.

\bibitem{gitfmm}
J.~Furtney {\it et al}.
\newblock {scikit-fmm.git}.
\newblock https://github.com/scikit-fmm/scikit-fmm.

\bibitem{gitadapQMMM}
Y.~Wang {\it et al}.
\newblock {AdapQMMM.jl.git}.
\newblock https://github.com/jameskermode/adaptive-qmmm-edge-disloc.git.

\bibitem{freund1998dynamic}
L.~Freund.
\newblock {\em Dynamic fracture mechanics}.
\newblock Cambridge university press, 1998.

\bibitem{gao02}
J.~Gao and D.~Truhlar.
\newblock Quantum mechanical methods for enzyme kinetics.
\newblock {\em Annu. Rev. Phys. Chem.}, 53:467--505, 2002.

\bibitem{Gavini:2007}
V.~Gavini, K.~Bhattacharya, and M.~Ortiz.
\newblock Quasi-continuum orbital-free density-functional theory: A route to
  multi-million atom non-periodic dft calculation.
\newblock {\em J. Mech. Phys. Solids}, 55(4):697--718, 2007.

\bibitem{Glukhova:2014}
O.~Glukhova, G.~Savostyanov, and M.~Slepchenkov.
\newblock A new approach to dynamical determination of the active zone in the
  framework of the hybrid model (quantum mechanics/ molecular mechanics).
\newblock {\em Procedia Materials Sci.}, 6:256--264, 2014.

\bibitem{grigorev2023calculation}
Petr Grigorev, Alexandra~M Goryaeva, Mihai-Cosmin Marinica, James~R Kermode,
  and Thomas~D Swinburne.
\newblock Calculation of dislocation binding to helium-vacancy defects in
  tungsten using hybrid ab initio-machine learning methods.
\newblock {\em Acta Materialia}, 247:118734, 2023.

\bibitem{herbst2021dftk}
M.~Herbst, A.~Levitt, and E.~Canc{\`e}s.
\newblock Dftk: A julian approach for simulating electrons in solids.
\newblock {\em Proceedings of the JuliaCon Conferences}, 3(26):69, 2021.

\bibitem{heyden2007}
A.~Heyden, H.~Lin, and D.~Truhlar.
\newblock Adaptive partitioning in combined quantum mechanical and molecular
  mechanical calculation of potential energy functions for multiscale
  simulations.
\newblock {\em J. Phys. Chem. B}, 111:2231--2241, 2007.

\bibitem{kerdcharoen1996}
T.~Kerdcharoen, K.~Liedl, and B.~Rode.
\newblock A {QM/MM} simulation method applied to the solution of {L}i$^+$ in
  liquid ammoia.
\newblock {\em Chem. phys.}, 211:313--323, 1996.

\bibitem{kermode08}
J.~Kermode, T.~Albaret, D.~Sherman, N.~Bernstein, P.~Gumbsch, M.~Payne,
  G.~Cs\'{a}nyi, and A.~De Vita.
\newblock Low-speed fracture instabilities in a brittle crystal.
\newblock {\em Nature}, 455:1224--1227, 2008.

\bibitem{lin2019mathematical}
L.~Lin and J.~Lu.
\newblock {\em A mathematical introduction to electronic structure theory}.
\newblock SIAM, 2019.

\bibitem{2020-pace1}
Y.~Lysogorskiy, C.~Oord, A.~Bochkarev, S.~Menon, M.~Rinaldi, T.~Hammerschmidt,
  M.~Mrovec, A.~Thompson, G.~Cs{\'a}nyi, C.~Ortner, et~al.
\newblock Performant implementation of the atomic cluster expansion (pace) and
  application to copper and silicon.
\newblock {\em Npj Comput. Mater.}, 7(1):1--12, 2021.

\bibitem{martin2020electronic}
R.~Martin.
\newblock {\em Electronic structure: basic theory and practical methods}.
\newblock Cambridge university press, 2020.

\bibitem{ogata01}
S.~Ogata, E.~Lidorikis, F.~Shimojo, A.~Nakano, P.~Vashishta, and R.~Kalia.
\newblock Hybrid finite-element/molecular-dynamic/electronic-density-functional
  approach to materials simulations on parallel computers.
\newblock {\em Comput. Phys. Commun.}, 138:143--154, 2001.

\bibitem{olson2019theoretical}
D.~Olson, C.~Ortner, Y.~Wang, and L.~Zhang.
\newblock Theoretical study of elastic far-field decay from dislocations in
  multilattices.
\newblock {\em ArXiv e-prints}, 1910.12269, 2019.

\bibitem{co2020}
C.~Ortner and J.~Thomas.
\newblock Point defects in tight binding models for insulators.
\newblock {\em Math. Models Methods Appl. Sci.}, 30:2753--2797, 2020.

\bibitem{2016-precon1}
D.~Packwood, J.~Kermode, L.~Mones, N.~Bernstein, J.~Woolley, N.~I.~M. Gould,
  C.~Ortner, and G.~Csanyi.
\newblock A universal preconditioner for simulating condensed phase materials.
\newblock {\em J. Chem. Phys.}, 144, 2016.

\bibitem{papaconstantopoulos15}
D.~Papaconstantopoulos.
\newblock {\em Handbook of the Band Structure of Elemental Solids, From Z = 1
  To Z = 112}.
\newblock Springer New York, 2015.

\bibitem{Ponga:2016}
M.~Ponga, K.~Bhattacharya, and M.Ortiz.
\newblock A sublinear-scaling approach to density-functional-theory analysis of
  crystal defects.
\newblock {\em J. Mech. Phys. Solids}, 95:530--556, 2016.

\bibitem{rice1968mathematical}
J.~Rice et~al.
\newblock Mathematical analysis in the mechanics of fracture.
\newblock {\em Fracture: an advanced treatise}, 2:191--311, 1968.

\bibitem{sethian1999fast}
J.~Sethian.
\newblock Fast marching methods.
\newblock {\em SIAM review}, 41(2):199--235, 1999.

\bibitem{Shapeev16}
A.~Shapeev.
\newblock Moment tensor potentials: {A} class of systematically improvable
  interatomic potentials.
\newblock {\em Multiscale Model. Simul.}, 14:1153--1173, 2016.

\bibitem{Slater54}
J.~Slater and G.~Koster.
\newblock Simplified {LCAO} method for the periodic potential problem.
\newblock {\em Phys. Rev.}, 94:1498--1524, 1954.

\bibitem{Verfurth:1996a}
R.~Verf\"{u}rth.
\newblock {\em A Review of A Posteriori Error Estimation and Adaptive
  Mesh-Refinement Techniques}.
\newblock John Wiley \& Sons Ltd., 1996.

\bibitem{waller2014}
M.~Waller, S.~Kumbhar, and J.~Yang.
\newblock A density-based adaptive quantum mechanical/molecular mechanical
  method.
\newblock {\em Chem. Phys. Chem.}, 15:3218--3225, 2014.

\bibitem{wang2020posteriori}
Y.~Wang, H.~Chen, M.~Liao, C.~Ortner, H.~Wang, and L.~Zhang.
\newblock A posteriori error estimates for adaptive qm/mm coupling methods.
\newblock {\em SIAM J. Sci. Comput.}, 43(4):A2785--A2808, 2021.

\bibitem{watanabe2014}
H.~Watanabe, T.~Kuba\v{r}, and M.~Elstner.
\newblock Size-consistent multipartitioning {QM/MM}: a stable and efficient
  adaptive {QM/MM} method.
\newblock {\em J. Chem. Theory Comput.}, 10:4242--4252, 2014.

\bibitem{wipf2007new}
David Wipf and Srikantan Nagarajan.
\newblock A new view of automatic relevance determination.
\newblock {\em Adv. Neural Inf. Process. Syst.}, 20, 2007.

\bibitem{witt2023acepotentials}
William~C Witt, Cas van~der Oord, Elena Gel{\v{z}}inyt{\.e}, Teemu
  J{\"a}rvinen, Andres Ross, James~P Darby, Cheuk~Hin Ho, William~J Baldwin,
  Matthias Sachs, James Kermode, et~al.
\newblock Acepotentials. jl: A julia implementation of the atomic cluster
  expansion.
\newblock {\em arXiv preprint arXiv:2309.03161, to appear in J. Chem. Phys.},
  2023.

\bibitem{ZYang2020}
Z.~Yang.
\newblock On-the-fly determination of active region centers in
  adaptive-partitioning {QM/MM}.
\newblock {\em Phys. Chem. Chem. Phys.}, 22(34):19307--19317, 2020.

\bibitem{zhang12}
X.~Zhang, Y.~Zhao, and G.~Lu.
\newblock Recent development in quantum mechanics/molecular mechanics modelling
  for materials.
\newblock {\em Int. J. Multiscale Comput. Eng.}, 10:65--82, 2012.

\bibitem{zhao2005fast}
H.~Zhao.
\newblock A fast sweeping method for eikonal equations.
\newblock {\em Math. Comput.}, 74(250):603--627, 2005.

\end{thebibliography}
